\documentclass[12pt,preprint]{aastex}
\usepackage{natbib}
\usepackage{amsmath}
\usepackage{graphicx}
\newcommand{\para}{\|}

\shorttitle{Particle Acceleration during MRI in an Accretion Disk}
\shortauthors{Hoshino}

\begin{document}

\title{Particle Acceleration during Magnetorotational Instability \\
       in a Collisionless Accretion Disk}

\author{Masahiro Hoshino}
\affil{Department of Earth and Planetary Science, The University of Tokyo, \\
       Tokyo, 113-0033. Japan}
\email{hoshino@eps.s.u-tokyo.ac.jp}

\begin{abstract}
Particle acceleration during the magnetorotational instability (MRI) in a collisionless accretion disk
was investigated by using a particle-in-cell (PIC) simulation.  We discuss the important role that magnetic reconnection 
plays not only on the saturation of MRI but also on the relativistic particle generation.
The plasma pressure anisotropy of $p_{\perp} > p_{\para}$ induced by the action of MRI dynamo leads
to rapid growth in magnetic reconnection, resulting in the fast generation of nonthermal particles with 
a hard power-law spectrum.  This efficient particle acceleration mechanism 
involved in a collisionless accretion disk may be a possible model to explain
the origin of high energy particles observed around massive black holes.
\end{abstract}


\section{Introduction}
One of the fundamental problems in astrophysics is to understand 
the outward transport of the angular momentum in an accretion disk 
in association with the mass accretion inward. 
The magnetorotational instability (MRI) has been investigated as an efficient 
mechanism for mass and angular momentum transport \citep{BalHaw91,BalHaw98}.  
It is found that a weakly magnetized disk with outwardly decreasing angular velocity generates magnetohydrodynamic
(MHD) turbulence, which can provide angular momentum transport at a greatly enhanced rate.
This mechanism is relevant to a large variety of high-energy astrophysical 
sources where gravitational energy is released through mass accretion.  

So far, many simulations in a framework of MHD approximation 
have been performed to understand quantitatively its angular momentum transport 
\citep[e.g.][]{Hawley91,Hawley92,Matsumoto95,Stone96,Sano04}.
The mean free path of plasma, however, is not necessarily smaller than the 
characteristic scale length for some classes of astrophysical accretion disks, 
and the kinetic behavior of MRI beyond the MHD approximation needs to be understood.  
For example, for an accretion disk around the super-massive black hole 
at the center of the Galaxy of Sagittarius A* (Sag A*), the accretion rate is believed to be
much less than the Eddington rate, and radiatively inefficient accretion flow (RIAF) 
models are discussed \citep[e.g.][]{Nara98,Quat03}. 
The accretion proceeds through a hot, low-density, collisionless plasma with 
a proton temperature higher than the electron temperature,
and the equipartition of plasma temperature is not realized during the 
accretion flow because of a low collision rate.  
In addition to the nonequilibrium temperature between protons and electrons, 
nonthermal high-energy particles are observed 
\citep[e.g.][]{Yuan03,Aharonian08,Chernyakova11,Kusunose12}.
It is believed that accretion disks around black holes or super massive black holes 
would be in the collisionless state of plasma.

Motivated by the observation of the collisionless accretion disk, 
\citet{Quat02,Sharma03,Sharma06} studied the linear instability of MRI 
by including both the effect of pressure anisotropy based on 
both the double adiabatic theory \citep{CGL56,Kulsrud83} 
and the kinetic effect of the Landau damping by using the
so-called ``Landau fluid approximation'' \citep{Hammett92}.
In a collisionless accretion disk, if the magnetic moment is conserved, the plasma pressure perpendicular 
to the magnetic field may increase in association with the stretching
magnetic field motion of MRI.  
In the linear theory,
\citet{Quat02,Sharma03} showed that the pressure anisotropy can
significantly modify the property of MRI and suggested that the
kinetic effect may be important for MRI saturation as well.

In addition to the study of the linear behavior, \citet{Sharma06} studied the
nonlinear evolution of MRI by using their extended MHD simulation model.
To incorporate the relaxation process of pressure anisotropy, 
they utilized the theoretical results of pressure anisotropy instabilities 
previously studied under homogeneous plasmas \citep[e.g.][]{Gary97} 
and introduced a model equation for the pressure isotropization process.
By using the model, they showed that in a collisionless plasma the rate of angular momentum transport 
can be moderately enhanced during the nonlinear stage.  

As their simulation study was based on a fluid model, it was important to 
investigate the kinetic effects in a fully kinetic approach.  
\citet{Riquelme12} performed two-dimensional PIC simulations
and confirmed the excitation 
of the mirror mode and relaxation of pressure anisotropy 
studied by the previous fluid-based model \citep{Sharma06}.  
In addition to the pressure anisotropy effect, 
the formation of a power-law energy spectrum during magnetic reconnection was pointed out.  
However, their particle-in-cell (PIC) simulations studied a low plasma 
$\beta = 8 \pi p /B^2$ regime of $\beta  = 0.05 \sim 40$ and used
a different boundary condition from the standard open shearing box 
technique \citep{Hawley95}.

In this work, we studied the collisionless MRI in a high plasma $\beta$ 
regime of $\beta = 96 \sim 6144$ with the standard open shearing box 
boundary condition.  
The pressure anisotropy generated in the course of
collisionless MRI is subject to not only the mirror mode and its associated
angular momentum transport but also the rapid growth of the
collisionless magnetic reconnection process.  It is known that 
the reconnection rate is strongly enhanced if the pressure 
perpendicular to the magnetic field is higher than the
parallel pressure.
The attention of the reader is drawn to the role of magnetic reconnection
with pressure anisotropy, which should lead to the production of significant nonthermal 
particle and the generation of MHD turbulence during collisionless MRI.

\section{PIC Simulation in Local Rotating Frame}
We focus on the local behavior of the collisionless 
MRI in the simplest representation of an accretion disk by 
using a PIC simulation code.  
Local simulations in the MHD framework have 
been performed by numerous researchers \citep{Hawley91,Hawley92,Matsumoto95,Stone96,Sano04}, 
and the generation of MHD turbulence 
and its angular momentum transport in an accretion disk also have been studied 
\citep{BalHaw98}.  The setup of our local kinetic simulation study 
is basically the same as those investigated in the MHD simulations, but we took 
into account plasma kinetic effects such as production of nonthermal particles
and plasma instabilities caused by pressure anisotropy during the evolution.

As shown by the earlier works
\citep{Sharma06,Riquelme12}, the basic behavior of the collisionless 
MRI is similar to that in an MHD framework; namely,
the amplification of a radial magnetic field by the stretching out of a vertical magnetic 
field and the formation of two inward- and outward-flowing streams.  
This stretching motion forms the channel flows with both a dense plasma gas and
a strong electric current, and then 
a violent magnetic field energy release process may 
be expected in the channel flow region by magnetic reconnection.  The magnetic reconnection
is known to have an important effect on the angular momentum transport
in association with the magnetic field amplification, but
our motivation in the kinetic simulation was not only to study the angular 
momentum transport but also to understand how and where the particle 
acceleration process occurs 
during MRI under collisionless magnetic reconnection.  

\subsection{Basic Equations}
In our kinetic study of MRI, we performed the PIC simulation in a local frame rotating 
with angular velocity $\Omega_0$ at a distance $r_0$ from the center. 
Because the magnetorotational instability is a class of the local instability,
this model was sufficient to capture much of the essential physics.
We solved a set of Maxwell's equations, which include the displacement current $\partial E/\partial t$.  
Because the local rotating frame with the angular velocity $\vec{\Omega}_0= \Omega_0 \vec{e}_z$ 
is not an inertial frame, some additional correction terms should be included in a set of 
Maxwell's equations \citep[e.g.][]{Schiff39}.  
The equations in the local rotating system after the Lorentz transformation can be written by
\begin{eqnarray}
  \frac{1}{c}\frac{\partial \vec{B}}{\partial t} &=& - \nabla \times \vec{E}, \\
  \nabla \cdot \vec{B} &=& 0, \\
  \frac{1}{c}\frac{\partial}{\partial t} \left( \vec{E} - \frac{\vec{v}_0}{c} \times \vec{B} \right) 
    &=& \nabla \times \vec{B}^* -\frac{4 \pi}{c} \vec{J}, \label{ampere_eq} \\
  \nabla \cdot \left( \vec{E}- \frac{\vec{v}_0}{c} \times \vec{B} \right) &=& 4 \pi \rho_c, \label{poisson_eq} 
\end{eqnarray}
where $\vec{v}_0(r)= \Omega_0 \vec{e}_z \times \vec{r}$ and
\begin{equation}
  \vec{B}^* = \vec{B} + \frac{\vec{v}_0}{c} \times \left(\vec{E}-\frac{\vec{v}_0}{c} \times \vec{B}\right).
\label{B_star}
\end{equation}
In these equations, we must assume that the distance of the local frame from the center is much less than the light cylinder 
($c/\Omega_0$), and we can neglect the terms that have the second order of $v_0/c$ in Eq.(\ref{B_star}). 
As a result, we assumed $\vec{B}^* = \vec{B}$.  

In the same approximation, the equations of motion including Coriolis, centrifugal, and gravity forces 
are given by
\begin{eqnarray}
  \frac{d\vec{p}}{dt} &=& e(\vec{E} + \frac{\vec{v}}{c} \times \vec{B})
  +m \gamma \left( 2 \vec{v} \times \vec{\Omega}_0 
  - \vec{\Omega}_0 \times (\vec{\Omega}_0 \times \vec{r}) - \frac{GM}{r^2} \vec{e}_r \right), \\
  \frac{d \vec{x}}{dt} &=& \vec{v},
\end{eqnarray}
where $\vec{p} = m \gamma \vec{v}$.  The vertical component of gravity is ignored.
We solved the above equations in Cartesian coordinates of $x$, $y$, and $z$, 
which correspond respectively to the radial direction $x=r-r_0$, 
the azimuthal direction $y=r_0 \phi$, and the parallel to the rotation axis $z$, 
where the position of $r_0$ is the center of the simulation box.
We also assumed that the size of the simulation box ($L'$) was much smaller than 
its distance to the center of the disk.
For a rotating disk around a central object with an angular velocity $\Omega(r)$, 
the force balance of $GM/r^2=r \Omega(r)^2$ is satisfied, 
and we use the tidal expansion of the effective potential with a constant 
$q=-\partial \rm{ln} \Omega/\partial \rm{ln} r$ at $r_0$, where 
the parameter $q$ is set to be $3/2$ for a Keplerian disk. 
The equation of motion becomes
\begin{eqnarray}
  \frac{d\vec{p}}{dt} &=& e(\vec{E} + \frac{\vec{v}}{c} \times \vec{B})
  -m \gamma(2 \vec{\Omega}_0 \times \vec{v}- 2q \Omega_0^2 x \vec{e}_x). 
\end{eqnarray}

In this work, we studied the time evolution in the meridional plane of $(x,z)$.
For simplicity, we set up positron $(e^+)$ and electron $(e^-)$ plasmas, 
and save the computational time in PIC code.  However, the linear dispersion relations of 
MRI in pair plasma is the same as in the standard MHD system (see Appendix A).  

The above basic equations include the velocity term of 
$\vec{v}_0(x,r_0) = \Omega_0(r_0) \vec{e}_z \times (x + r_0) \vec{e}_x$, which is a function of
the distance $r_0$ from the center, and this term arising from the effect of a noninertial frame 
appears only as the correction of the displacement current in the Amp\`ere equation 
and of the electric charge density in the Poisson equation, so that those terms can be,
in general, neglected in a low-frequency and long-wavelength MHD regime.  In a high- frequency regime,
where the amplitude of the electric field is of the same order as that of the magnetic field, i.e.,$O(E) \sim O(B)$, 
the term $v_0 B/c$ is smaller than $E$ if $v_0 \ll c$.
For instance, the correction of the noninertial frame term appears as the
split of the free-radiation mode.  Namely, the phase velocity of the light wave $v_{ph}$ becomes 
$v_{ph}/c= \pm \sqrt{1 + (\vec{k} \cdot \vec{v}_0)^2/(2 k c)^2}
-( \vec{k} \cdot \vec{v}_0)/(2 k c)$, where $k$ is the wave number.  
We find that the correction is very small as long as $v_0 \ll c$ again.
To check the above statement, we performed several simulation runs 
and compared the results with and without the noninertial frame term,
and we found that the results do not change much as long as $L'/2 \le r_0 < 3 L'/2$, where
$L'$ is the size of the simulation box.
It is not easy to set up the simulation parameter satisfying 
both $r_0 \gg L'$ and $\Omega_0 r_0 \ll c$ in PIC code, because the grid size
is limited by the Debye length.
As we are interested in the nonlinear evolution of MRI in the MHD frequency range, 
we may neglect the noninertial frame term with $\vec{v}_0 \times \vec{B}/c$ 
in a set of Maxwell's equations. In this paper, we have assumed $\vec{v}_0=0$.  

\subsection{Initial and Boundary Conditions}
At the initial condition, a drift Maxwellian velocity distribution function was assumed
for both electrons and positrons in the local rotating frame with the angular velocity 
$\Omega_0(r_0)$. The drift velocity in the $y$-direction $v_y(x)$ was given by
\begin{equation}
  v_y(x)=r \Omega(r) - r \Omega_0(r_0) 
  \simeq \frac{\partial \rm{ln} \Omega}{\partial \rm{ln}r}\Big|_{r_0} \Omega_0(r_0)x 
  = -q \Omega_0(r_0)x,
\end{equation}
and other components were $v_x = v_z = 0$.  

A nonrelativistic, isotropic plasma pressure with a high plasma 
$\beta = 8 \pi (p_+ + p_-)/B_0^2$ was assumed, 
where the electron and positron gas pressures were related to the thermal velocities 
$v_{t\pm}$ by $p_\pm = (3/2) m_\pm n v_{t\pm}^2$.  
The initial magnetic field was set to be purely vertical to the accretion disk; i.e., 
$\vec{B}=(0,0,B_z)$ in the two-dimensional $x-z$ plane. 
The plasma parameters used in this paper are listed in Table \ref{tbl1}. 
The ratio of cyclotron frequency to the disk angular velocity was fixed to be 
$\Omega_{c\pm}/\Omega_0= \pm 10$, where $\Omega_{c\pm}=e_\pm B_0/m_\pm c$.
The grid size $\Delta$ was set to be $ 2^{3/2}(v_{t\pm}/\Omega_{p\pm})$, 
where $\Omega_{\pm}=\sqrt{8 \pi n e^2/m_\pm}$ is the pair plasma frequency.
The Alfv\'en velocity is defined by $V_A=B/\sqrt{8 \pi m_{\pm} n}$, so that
the plasma $\beta$ is equal to $3 v_{t\pm}^2/V_A^2$.  $N_x$ and $N_z$ are the grid size 
of $x$ and $z$ directions, and we assumed $N_x = N_z$ in this paper. 
$L_x=L_z=(N_x \Delta)/\lambda$ is the physical size normalized by $\lambda=2 \pi V_A.\Omega_0$.
$N_p$/cell is the number of 
particles per cell, and a large number of particles is necessary for
a high plasma $\beta$ simulation to suppress electrostatic fluctuations.

The uniform magnetic field in space is a simple assumption, but 
the uniform magnetic field $B_z$ is not an exact equilibrium solution 
because of a finite charge separation in a kinetic regime.  
Although the charge separation was very small,
we added a small correction term to the uniform magnetic field $B_0$,
and we used the initial condition of
\begin{equation}
  B_z(x)=B_0/\sqrt{1- \alpha' x^2},
\end{equation}
where $\alpha' = (q \Omega_0/c)^2$ and $\alpha' L'^2 \ll 1$, where $L'=N_x \Delta$ is the physical
system size.  The details are given in Appendix B.  

As the boundary condition, we employed the so-called shearing box approximation \citep{Hawley95}, 
where the differential rotation velocity $r \Omega_0(r_0)$ is calibrated for the $x$ boundary condition, 
and the physical quantities at the boundary are obtained by the Lorentz transformation 
according to the velocity difference of the Keplerian motion between the inner edge and the outer edge.
The physical quantities between the inner edge at $x_i=-(L_x \lambda)/2$ and 
the outer edge at $x_o=+(L_x \lambda)/2$ can be related by
\begin{eqnarray}
  \vec{E}(x_i,y^*,z) &=& \vec{E}(x_o,y,z)+\frac{1}{c} \vec{v}_{*} \times \vec{B}(x_o,y,z), \\
  \vec{B}(x_i,y^*,z) &=& \vec{B}(x_o,y,z), \\
  n_{\pm}(x_i,y^*,z) &=& n_{\pm}(x_o,y,z), \\
  \vec{J}_{\pm}(x_i,y^*,z) &=& \vec{J}_{\pm}(x_o,y,z) + e \vec{v}_{*} n_{\pm}(x_o,y,z), 
\end{eqnarray}
where $y^*=y+q \Omega_0 (L_x \lambda) t$ and $\vec{v}_{*} = q \Omega_0 (L_x \lambda) \vec{e}_y$, 
and we assumed $v_{*}^2/c^2 \ll 1$ and $\gamma_{*}=1/\sqrt{1-v_{*}^2/c^2} \sim 1$.
The total electric current $\vec{J}=\vec{J}_+ + \vec{J}_-$ can be obtained after 
the above Lorentz transformation of the pair plasma densities $n_{\pm}$ at the boundary.  
The double periodic boundary condition was used for both $y$ and $z$ directions.
In our three-dimensional code, we checked that the total magnetic field components were 
conserved during the time evolution; i.e.,
$\int_V \vec{B} dx dy dz = \rm{const}$.

In the previous kinetic simulation reported by \citet{Riquelme12}, they avoided the Lorentz transformation 
of the electric current at the boundary and implemented shearing coordinates, in which each grid moves 
with the shearing velocity $v = -q \Omega_0 x$.  Our simulation did not use the shearing coordinates and
adopted the same boundary condition used in MHD simulations \citep{Hawley95}.

To solve the kinetic MRI phenomena with better numerical stability, 
we slightly modified the simulation scheme used in our previous simulation studies.
The algorithm is briefly described in Appendix C.  Although we have developed a three-dimensional code 
in the shearing box approximation, in this paper, we only discuss the two-dimensional, meridional evolution of MRI
in the $x-z$ plane.

\section{Simulation Study of MRI in Meridional Plane}
In this section, we show the time evolution of kinetic MRI by paying special attention 
to several ingredients such as particle acceleration, magnetic reconnection, and pressure anisotropy.  
MRI is known to possess a dynamo action 
by stretching the magnetic field lines under differential rotation motion in a disk, 
and the antiparallel magnetic field induced by the dynamo action forms the current sheet.
If the magnetic diffusion process exists in the current sheet, it is expected that magnetic reconnection occurs.
Therefore, the balance between the dynamo action and the magnetic field-dissipation process 
by magnetic reconnection may control the saturation of magnetic field evolution.  
It is argued that the plasma $\beta = 8 \pi p/B^2$ is of the order of $10 \sim 1$ in
the saturation stage, and the equipartition of energy can be realized \citep{Hawley95,Stone96,Sano01}.
During the stage of magnetic reconnection, particle acceleration is highly expected,
because it is known that the collisionless magnetic reconnection can
quickly produce nonthermal particles with energies that exceed the thermal energy
\citep{Zenitani01,Zenitani05}.  
It is interesting to understand the behavior of particle acceleration during MRI.

\subsection{From Linear to Nonlinear Evolution}
\begin{figure}
\includegraphics[scale=0.65]{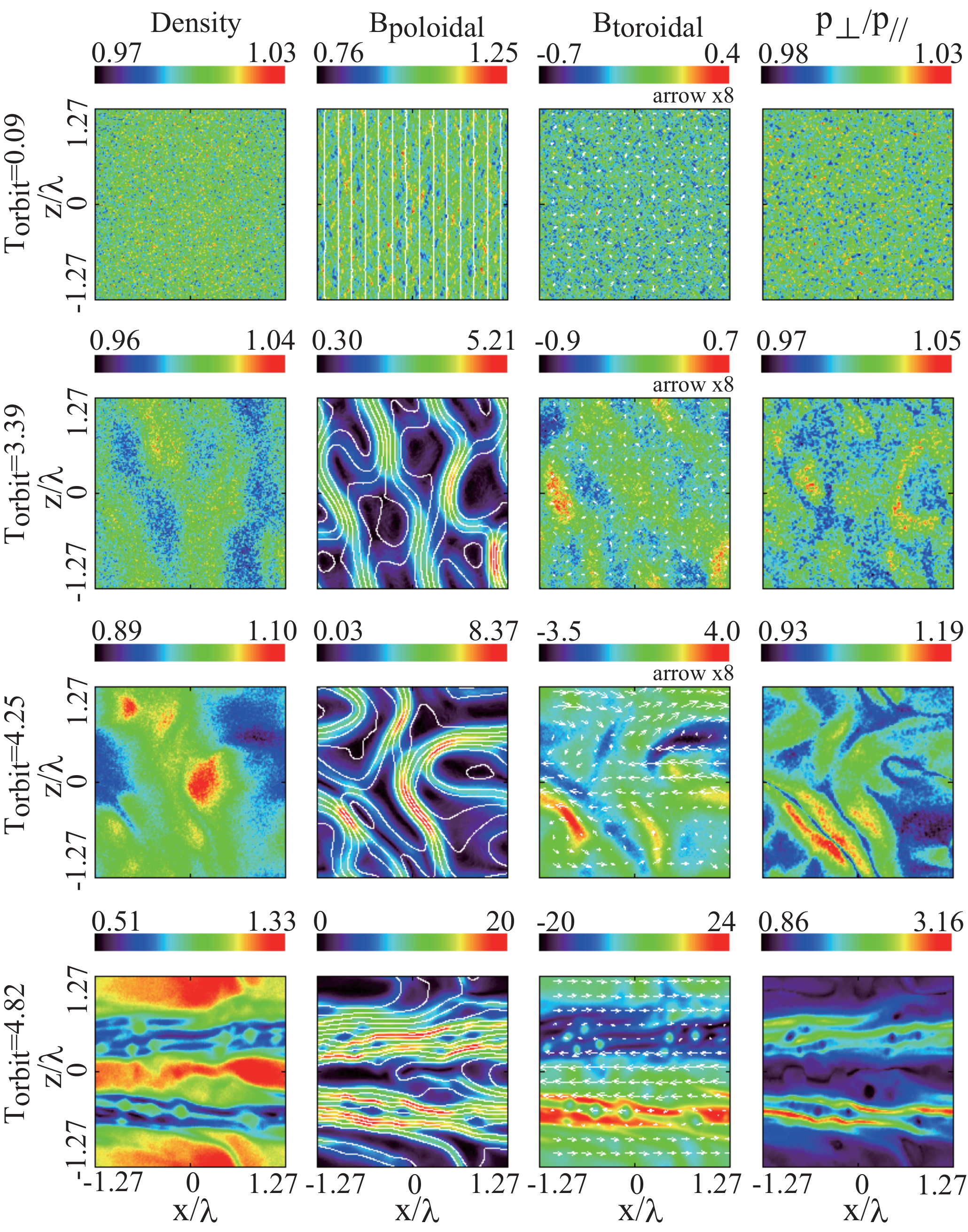}
\caption{Early time evolution of magnetorotational instability (MRI) for plasma $\beta=1536$.  
From top to bottom, the initial stage with the uniform plasma and the isotopic pressure (top), 
the linear growing stages under the coupling between MRI and mirror modes (2nd and 3rd columns), 
and the formation of channel flow (bottom).  From left to right, the pair-plasma density (left), 
the intensity of poloidal magnetic field $\sqrt{B_x^2+B_z^2}$ and the projection of magnetic field 
lines denoted by the white lines (2nd row), the toroidal magnetic field $B_y$ and the projection of 
the plasma flows by the white arrows (3rd row), and the anisotropy of the pair-plasma 
pressure $p_{\perp}/p_{\para}$ (right).  The density and magnetic field are normalized by the initial state.  
The magnitudes are shown in each top panel by the color bar with the linear scale.
The $x$ and $z$ coordinates are normalized by $\lambda = 2 \pi V_A/\Omega_0$.
\label{fig1}}
\end{figure}

Figure \ref{fig1}, from top to bottom, shows the time evolution of MRI for $\beta=1536$ 
from the initial state to the nonlinear stage.
The top four panels are the initial stage with the orbital time of 
$T_{\rm orbit} = t \Omega_0/2 \pi = 0.096$, 
and the other panels from the second row to the fourth row correspond to $T_{\rm orbit}=3.39$, $4.25$ and $4.82$. 
In each row, the pair plasma density ($n=(n_+ + n_-)/2$ defined by both electron and positron components), 
the poloidal magnetic field $(B_p=\sqrt{B_x^2+B_z^2})$, the toroidal magnetic field $(B_{y})$, 
and the pair pressure anisotropy $(p_\perp/p_\para)$ defined by the pair components
are depicted from left to right.
The white lines in the meridional field show the magnetic field lines obtained from 
the contour of the vector potential of $A_{y}(x,z)$, and the white arrows superposed 
in the toroidal magnetic field are the pair plasma flow vectors 
$(\vec{v}_- + \vec{v}_+)/2$ projected onto the plane.
The plasma density and the magnetic fields were normalized by the initial density and
the initial total magnetic field, respectively.  
The $x$ and $z$ coordinates were normalized by $\lambda = 2 \pi V_A/\Omega_0$.

At $T_{\rm orbit}=0.096$ (1st row), the plasmas and magnetic field still remained
almost in the initial state, and the structures were uniform in space except for some fluctuations.  
At $T_{\rm orbit}=3.39$ of the linear growth stage (2nd row), 
the magnetic field lines are largely folded, and one can see the periodic variation of
the poloidal magnetic field intensity in both $x$ and $z$ directions, suggesting that
the oblique propagating modes were excited.  
The magnitude $B_p$ in the intensified region was about five times larger than the initial value.
Variations in plasma density, toroidal magnetic field and pressure anisotropy
could also be observed, but the magnitudes of such variations were less than about $10 \%$,
 suggesting that the unstable mode is basically incompressible 
and that MRI is mainly provided by the transverse Alfv\'enic/slow mode wave in high $\beta$ plasma.
With time, the magnetic field lines were further folded at $T_{\rm orbit}=4.25$, and
at $T_{\rm orbit}=4.82$, a pair of stratified plasma structures associated with 
counter streaming flows can be observed. The fast flow regions correspond to the high-density 
current sheet where the toroidal magnetic field polarity changes.  
These properties are similar to the standard MRI time evolution in the MHD framework.

One of the important agents in the kinetic MRI mode is the generation of pressure anisotropy
\citep{Quat02,Sharma03,Sharma06}.
We can see the evolution of pressure anisotropy with $p_{\perp} > p_{\para}$ (4th column),
where $p_{\perp}$ and $p_{\para}$ are the plasma pressure perpendicular and parallel to the magnetic field, 
 respectively.
Roughly speaking, the pressure anisotropy has a good correlation with the total magnetic field.
At $T_{\rm orbit}=4.82$, the strong anisotropy can be clearly seen in the strong magnetic field regions,
which correspond to the reddish regions in the poloidal magnetic field and the blueish and reddish regions
in the toroidal magnetic field.  
At $T_{\rm orbit}=4.82$, the anisotropy reaches $p_{\perp}/p_{\para} \sim 3.16$

At $T_{\rm orbit}=3.39$ of the early linear growth stage, the plasma density has a good correlation
with the total magnetic field because of the plasma property of the so-called frozen-in condition.
At $T_{\rm orbit}=4.25$, however, the correlation between them is not necessarily good, and
at $T_{\rm orbit}=4.82$, the plasma density has an anticorrelation against the total magnetic field,
which may suggest the slow mode wave behavior.

The formation of the strong pressure anisotropy can be easily understood by 
a double adiabatic equation of state \citep{CGL56,Kulsrud83}, and 
this behavior has already been discussed in previous kinetic simulations \citep{Sharma06,Riquelme12}.
In collisionless plasma, the equation of state describing the evolution of 
$p_{\para}$ and $p_{\perp}$ can be given by
\begin{eqnarray}
\frac{D}{Dt} \left( \frac{p_{\perp}}{\rho B} \right) &=& 0, \label{CGL1} \\
\frac{D}{Dt} \left( \frac{p_{\para} B^2}{\rho^3} \right) &=& 0, \label{CGL2}
\end{eqnarray}
where the heat flux transport is neglected for simplicity.
From the above first equation, which shows the conservation of the first adiabatic motion of the particle, 
we can understand the enhancement in perpendicular pressure $p_{\perp}$ during the time evolution of MRI.  
The total magnetic field increases during MRI, and the plasma density can also increase by the nature of
the frozen-in behavior, but the enhancement in plasma density is relatively small
in a high plasma $\beta$ medium.  Then, the perpendicular pressure will have to increase with an increasing 
total magnetic field.  On the other hand, as we can find from the second equation 
that can be obtained from the first and second adiabatic motions of particle,
we know that the parallel pressure decreases slightly with an increasing total magnetic field.
In fact, we observed the above behavior in our simulation results.
Therefore, it is a natural consequence that the strong pressure anisotropy is induced 
during the evolution of MRI under the stretching motion of a magnetic field 
because of the differential rotation of a Keplerian disk; 
i.e., the MRI dynamo action in association with the folded magnetic field.

Because the pressure anisotropy has free energy in the system, isotropization of plasmas can occur
through some plasma instabilities.  Under the anisotropy pressure with $p_{\perp}/p_{\para} > 1$ 
in a high $\beta$ plasma, the mirror-mode instability and the ion-cyclotron instability 
may have important effects on the isotropization \citep{Gary94,Gary97}.  
As discussed by \citet{Sharma06}, if we use the criteria of the onsets of 
pitch-angle scattering by the mirror mode and the ion-cyclotron instabilities of
$p_{\perp}/p_{\para} -1 > 7/\beta_{\perp}$ 
and $p_{\perp}/p_{\para} -1 > 0.35/\beta_{\para}^{0.42}$, respectively,
we find that the intersection point of two lines is 
$(\beta_{\para}, p_{\perp}/p_{\para} -1) \sim (187, 0.039)$, and that
for a high plasma $\beta_{\perp} > \beta_{\para} > 187$,
the onset of the mirror mode occurs faster than the ion-cyclotron instability 
during the increase in pressure anisotropy.

Because the mirror mode is generated via the instability of the slow MHD wave, 
the excitation of the mirror mode can be inferred from the anticorrelation 
between the magnetic pressure and the gas pressure.  For instance, at $T_{\rm orbit}=4.82$, 
we can see several small spots in the strong anisotropic pressure regions, and 
we found that the plasma density was high in those regions, 
whereas the magnitude of the magnetic field and the pressure anisotropy were reduced.
The behavior is regarded as the mirror-mode property with the diamagnetic effect.

The size of these spots is of the same order as the gyro-radius for the heated plasma 
under the amplified magnetic field, because the initial gyro-radius was $56.6 \Delta = 0.36 \lambda$, 
and the local magnetic field is enhanced several tens of times over the initial value
at $T_{\rm orbit}=4.82$.
This is also consistent with the kinetic linear theory of the mirror mode \citep{Gary94,Gary97}.
In this way, we observe the excitation of the mirror-mode instability,
and the pressure anisotropy can be reduced through the excitation of the mirror-mode wave.

\begin{figure}
\includegraphics[scale=0.65]{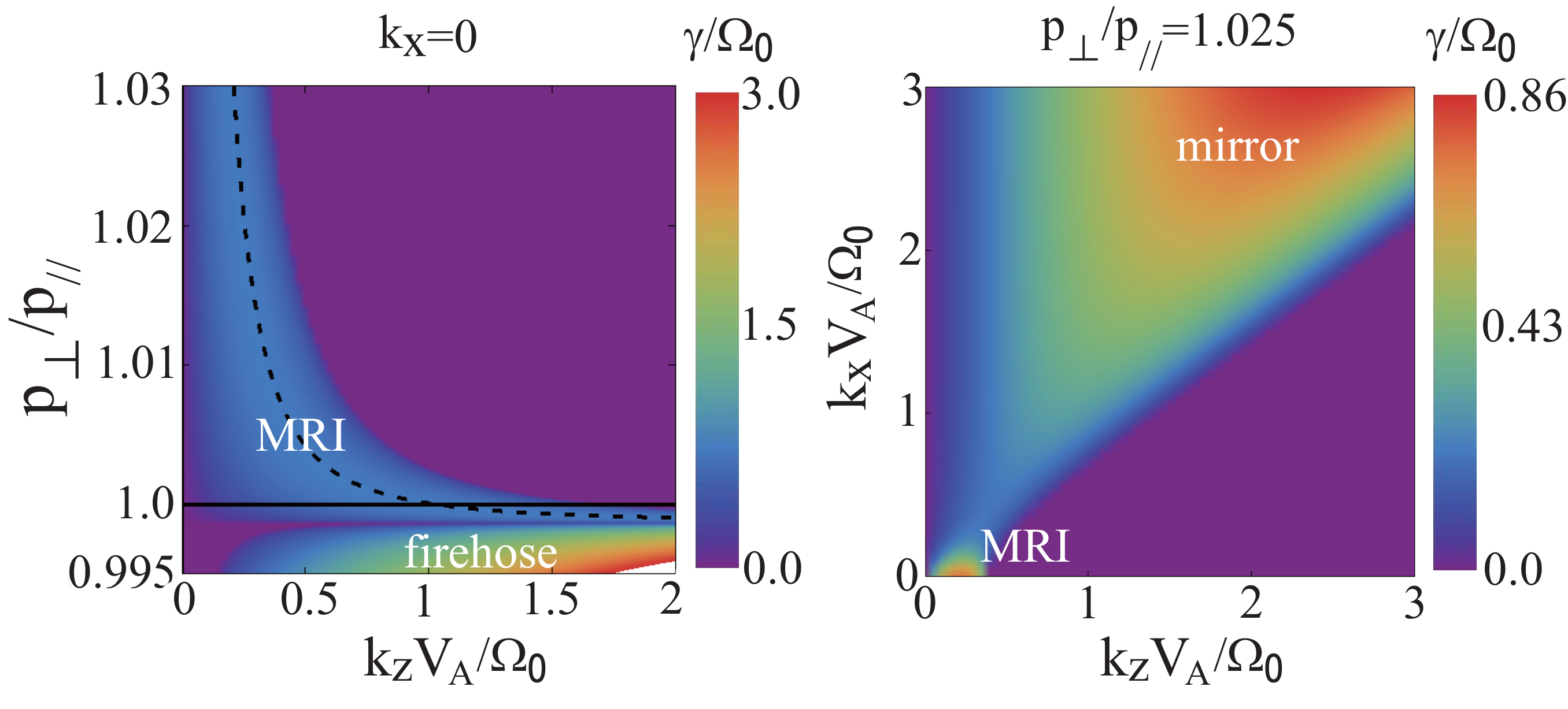}
\caption{Normalized linear growth rate $\gamma/\Omega_0$ for MRI, mirror and fire-hose modes 
for the plasma $\beta=1536$.  The left-hand panel shows the linear growth rate 
as the function of the wave number $k_z$ parallel to the initial magnetic field (i.e., the rotation axis $\Omega_0$) 
and pressure anisotropy $p_{\perp}/p_{\para}$. The perpendicular wave number $k_x=0$ is assumed.  
The black dashed lines denote the maximum growth rate for MRI predicted by theory.  
The fire-hose unstable region appears when $p_{\perp}/p_{\para} < 1$.  
The right-hand panel shows the growth rate as the function of $k_z$ and $k_x$ 
with $p_{\perp}/p_{\para}=1.025$.  The MRI mode is located in the small parallel wave number region, 
whereas the unstable mirror mode appears in the large wave number region with the oblique propagation.
\label{b=1536_theory}}
\end{figure}

It is interesting to note that the finite pressure anisotropy with $p_{\perp}/p_{\para} > 1$ 
leads to shifting the unstable mode of the family of MRI to the longer-wavelength region \citep{Sharma06}.  
The left-hand panel in Figure \ref{b=1536_theory} shows the linear growth rates of
the MRI modes as a function of $k_z$ and $p_{\perp}/p_{\para}$ with the same 
theoretical framework as studied by \citet{Quat02,Sharma06}.  For a better description of
the mirror mode, the heat flux effect is taken into account by using 
the so-called Landau fluid closure \citep{Hammett92,Sharma03}.  
The most unstable mode for the case of the isotropic plasma with $p_{\perp}/p_{\para} = 1$
is located around $k V_A/\Omega_0 \sim 1$, and the unstable mode shifts toward the smaller $k_z$
as increasing $p_{\perp}/p_{\para}$.  
If the value of $p_{\perp}/p_{\para}$ is slightly smaller than unity, the growth rate is
quickly reduced, but the fire-hose instability can be excited in a short-wavelength region.

The dashed curve shows the position of the maximum growth rate obtained from 
the linear theory in a high plasma $\beta$ approximation, which can be expressed by
\begin{equation}
  (k_z V_A)^2 \left(1-\frac{\beta_{\para}-\beta_{\perp}}{2} \right)=\frac{15}{16}\Omega_0^2.
\end{equation}  
For the parallel propagation mode, the effect of pressure anisotropy appears only through
the Alfv\'en speed, because there is no coupling between the transverse mode and 
the compressional/longitudinal wave.  The magnetic tension force $f_{\rm tension}$ 
can be modified as $f_{\rm tension} = (k_z V_A)^2 (1-(\beta_{\para}-\beta_{\perp})/2)$,
and the Alfv\'en speed is also modified according to the tension force.
On the other hand, the corresponding growth rate can be given by
\begin{equation}
 {\rm Im}(\omega) = \frac{3}{4}\Omega_0,
\end{equation}
which is the same as in the case of isotropic plasma \citep{BalHaw98}.

\begin{figure}
\includegraphics{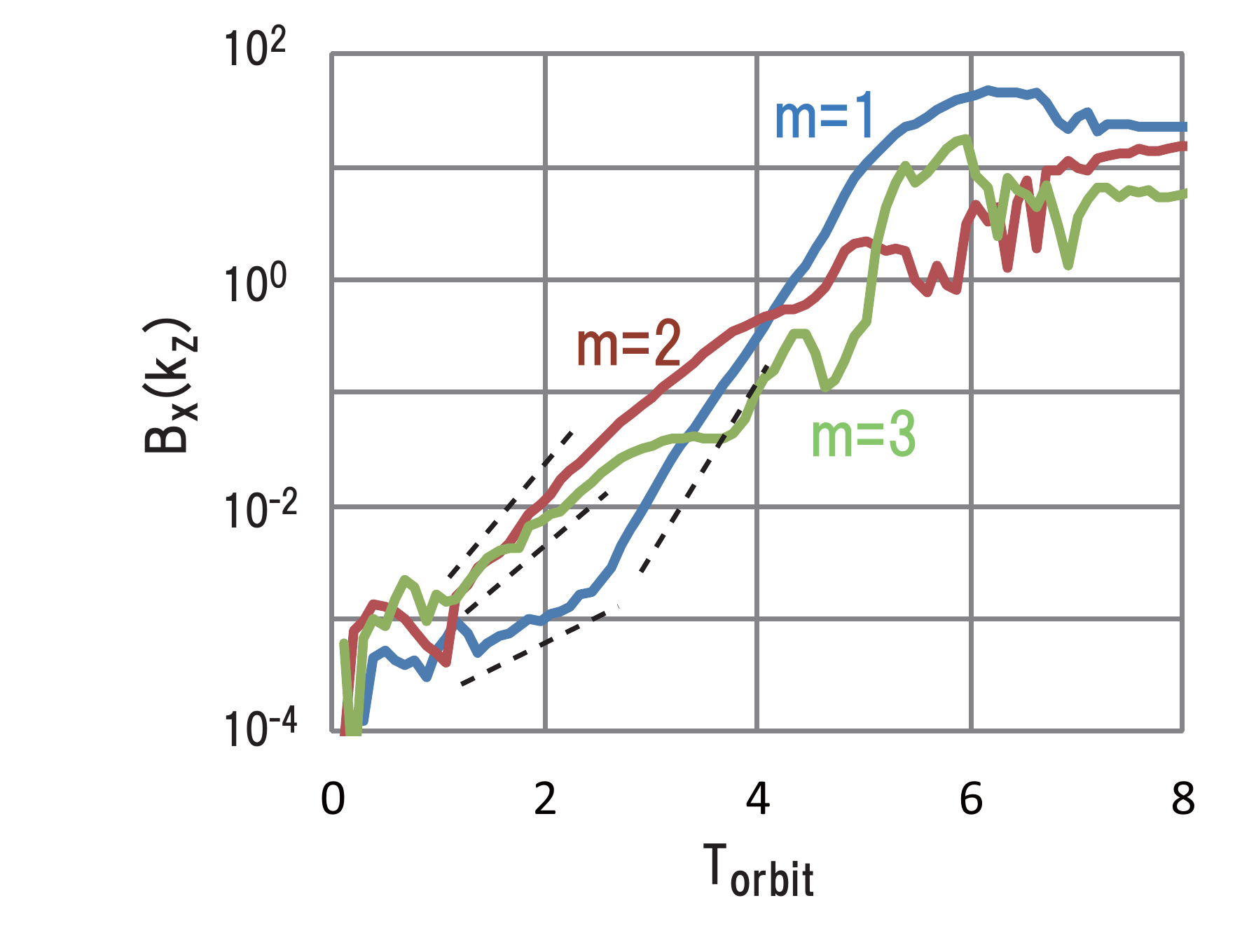}
\caption{Time history of Fourier modes for $B_x(k_z)$ propagating parallel to the rotation axis $\Omega_0$.  
During the early linear growth stage from $T_{\rm orbit}=1 \sim 2.5$, 
the mode number $m=1$ is the fastest growing wave.  After $T_{\rm orbit}>2.5$, the longer wavelength 
mode with $m=1$ can grow faster.  After $T_{\rm orbit} \sim 4$, the mode dominates the system and forms the channel flow.
\label{fft_mode}}
\end{figure}

Let us revisit the simulation result.  
During the early evolution of MRI, we observed the simultaneous onset of mirror mode and the isotropization of 
 pressure anisotropy, but the pressure isotropization was not completed,
and some finite pressure anisotropy with $p_{\perp}/p_{\para} > 1$ may remain. In fact, 
from another analysis, we find that the volume-averaged pressure anisotropy 
$<p_{\perp}/p_{\para}>$ is almost isotropic until $T_{\rm orbit}=3$, but 
$<p_{\perp}/p_{\para}>$ gradually increases up to $1.01$ at $T_{\rm orbit}=4$, 
and after $T_{\rm orbit}=4$, $<p_{\perp}/p_{\para}>$ dramatically increases (not shown here).
Therefore, it is highly probably that the unstable MRI mode may shift toward 
a longer-wavelength range during the evolution of pressure anisotropy.
Figure \ref{fft_mode} shows the time history of Fourier modes of $k_z V_A/\Omega_0=0.39$, $0.78$, and $1.18$,
which correspond to the mode number of $m=1$, $2$, and $3$, where $k_z=(m/N)(2 \pi/\Delta)$,
where $N$ is the number of the grid point, $\Delta$ is the grid size, and $N \Delta = 2.55 (2 \pi V_A/\Omega_0)$
in our simulation parameter (see Table \ref{tbl1}).
Note that the very early stage of $T_{\rm orbit}=0 \sim 1$ is probably a preinstability stage under the thermal
fluctuation, and the evolution of this time interval may depend on the level of the initial thermal fluctuation.  
After the preinstability stage, one can find that the mode of $m=2$ grows faster than others 
until $T_{\rm orbit}<2.5$, but after $T_{\rm orbit}>2.5$, the mode of $m=1$ grows rapidly, 
and eventually the longest-wavelength mode of $m=1$ dominates the system, 
which corresponds to the formation of a pair of the channel flows in Figure \ref{fig1}.
The linear growth rates $\gamma/\Omega_0$ can be measured from the slopes of the growth curves
indicated by the dashed lines in Figure \ref{fft_mode},
and $\gamma/\Omega_0$ of $m=1$, $m=2$, and $m=3$ before $T_{\rm orbit}<2.5$ are 
$0.18$, $0.48$, and $0.30$, respectively. $\gamma/\Omega_0$ of $m=1$ after $T_{\rm orbit}>2.5$ is $0.55$.
These values are smaller than the theoretical maximum growth rate with $\gamma/\Omega_0=0.75$,
because the evolution of time-dependent pressure anisotropy may modify the linear behavior.

In addition to the unstable mode shift towards the longer wavelength, the generation of
oblique propagating modes is also important in the kinetic MRI evolution.
The right-hand panel in Figure \ref{b=1536_theory} shows the linear growth rates 
as a function of $k_z$ and $k_x$.  We fix $p_{\perp}/p_{\para}=1.025$.
We find that the MRI mode is localized around $k_z V_A/\Omega_0 \sim 1/4$, while 
the mirror mode appears for obliquely propagating waves with large wave numbers.
(Note that this is the fluid-based model, where the growth rate of the mirror mode
increases with increasing wave number.  The kinetic effects, such as the finite Larmor 
radius effect should be taken into account for the understanding of the suppression of 
the mirror mode in the large wave number regime.)
In addition to the MRI and mirror modes, we can see a coupled mode regime 
connecting to two unstable modes \citep{Quat02,Sharma03,Sharma06}.

The oblique mode seen at $T_{\rm orbit}=3.39$ in Figure \ref{fig1} can be interpreted 
by the coupling and the mirror modes. The simulation does not start from a finite 
pressure anisotropy, but 
the pressure anisotropy is generated during the evolution.  As the magnitude of the 
pressure anisotropy is not uniform in space, it is not necessarily good to
compare the result of Figure \ref{b=1536_theory} with the simulation result.
However, the effect of pressure anisotropy is very strong in a high plasma $\beta$,
and we think that the effect should appear even for the early evolution phase of MRI.
\citet{Sharma06} found that the saturation of MRI remains at a very low level
if the pitch-angle scattering model is not incorporated in their extended MHD simulation, 
because the pressure anisotropy suppresses the evolution of MRI.
Then, \citet{Sharma06} implemented the pitch-angle scattering model into their simulation
and demonstrated that the saturation level that is almost same as 
the previous MHD simulation results can be recovered.
In our PIC simulation study, the pitch-angle scattering process was included self-consistently,
but the ratio of the cyclotron frequency $\Omega_c$ to the rotation angular frequency $\Omega_0$ 
was set to be $\Omega_c/\Omega_0=10$, 
which should have a much larger value in a realistic situation.  
Therefore, the level of the pressure anisotropy discussed in our simulation would be larger than
that expected in a realistic situation. Nevertheless, we observed that the nonlinear evolution 
of MRI was possible in the self-consistent simulation.

\subsection{Onset of Magnetic Reconnection under Pressure Anisotropy}
As time goes on, the current sheet at the interface of the antiparallel magnetic field 
in the channel flow becomes thinner, 
and the onset of magnetic reconnection can be expected in these thin current sheets.  
Although channel flow is an exact nonlinear solution of MRI without diffusivity,
the possibility of the parasitic instability after the stage of channel flow 
has been discussed by \citet{Goodman94}.  It is interesting to study the nonlinear evolution
in a kinetic system, where collisionless/inertia resistivity is self-consistently included. 

\begin{figure}
\includegraphics[scale=0.65]{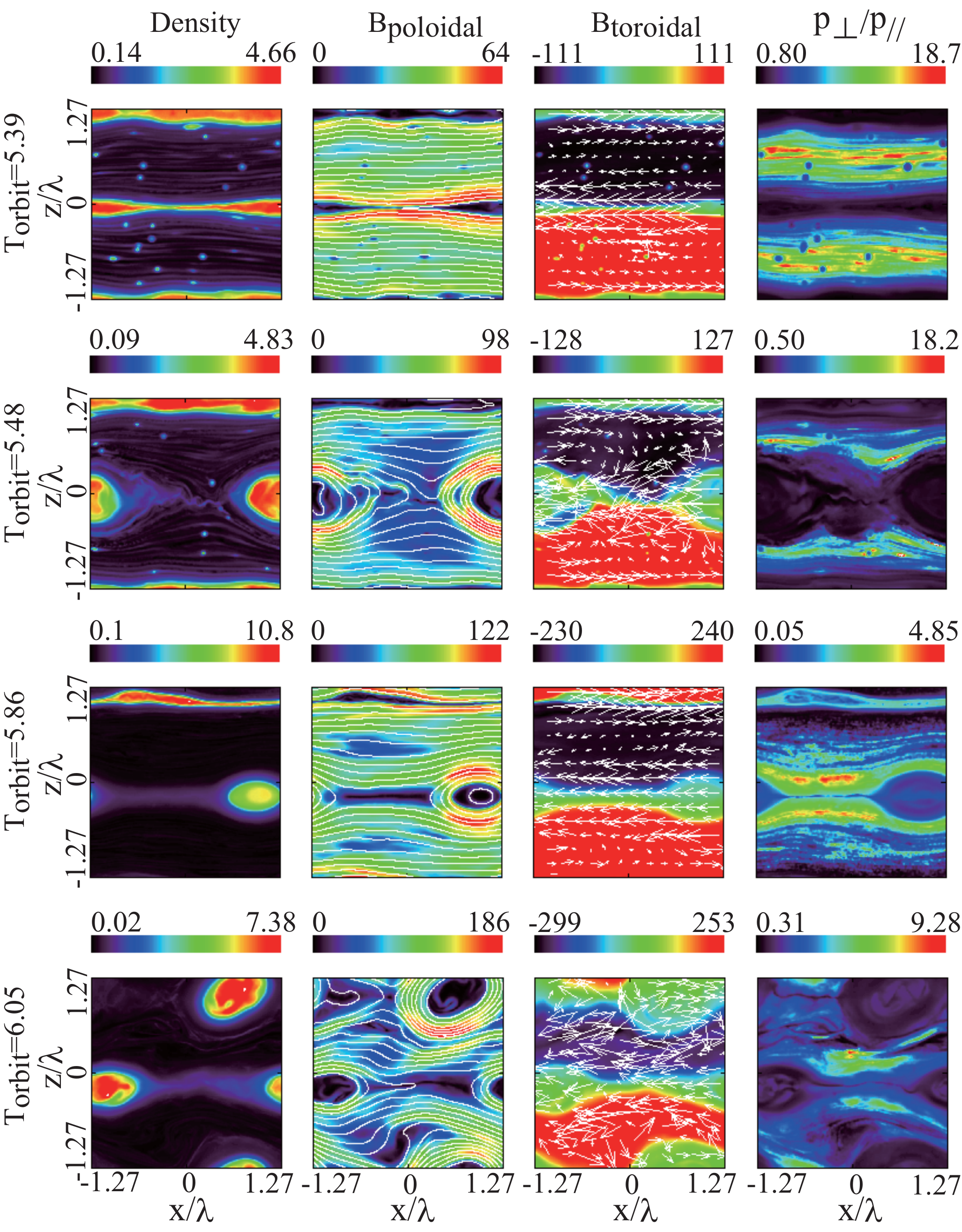}
\caption{Evolution of thin current sheets and magnetic reconnection of $T_{\rm orbit}=5.39 \sim 6.05$ after the formation of channel flow.  The format is same as that of Figure \ref{fig1}.
\label{fig2}}
\end{figure}

The nonlinear evolution after the formation of a thin current sheet is shown in Figure \ref{fig2}.
The formation of the X-type neutral line was observed at $T_{\rm orbit}=5.39$, 
and the disruption of the current sheet occurred at $T_{\rm orbit}=5.48$.  
For another current sheet situated around $z/\lambda \sim 1.2$, the thinning of
the current sheet was observed at $T_{\rm orbit}=5.86$, and 
rapid disruption of the current sheet occurred at $T_{\rm orbit}=6.05$.
We observed the dynamic evolution of magnetic reconnection associated with
the thinning of the current sheet after the formation of the channel flows.

\begin{figure}
\includegraphics[scale=0.65]{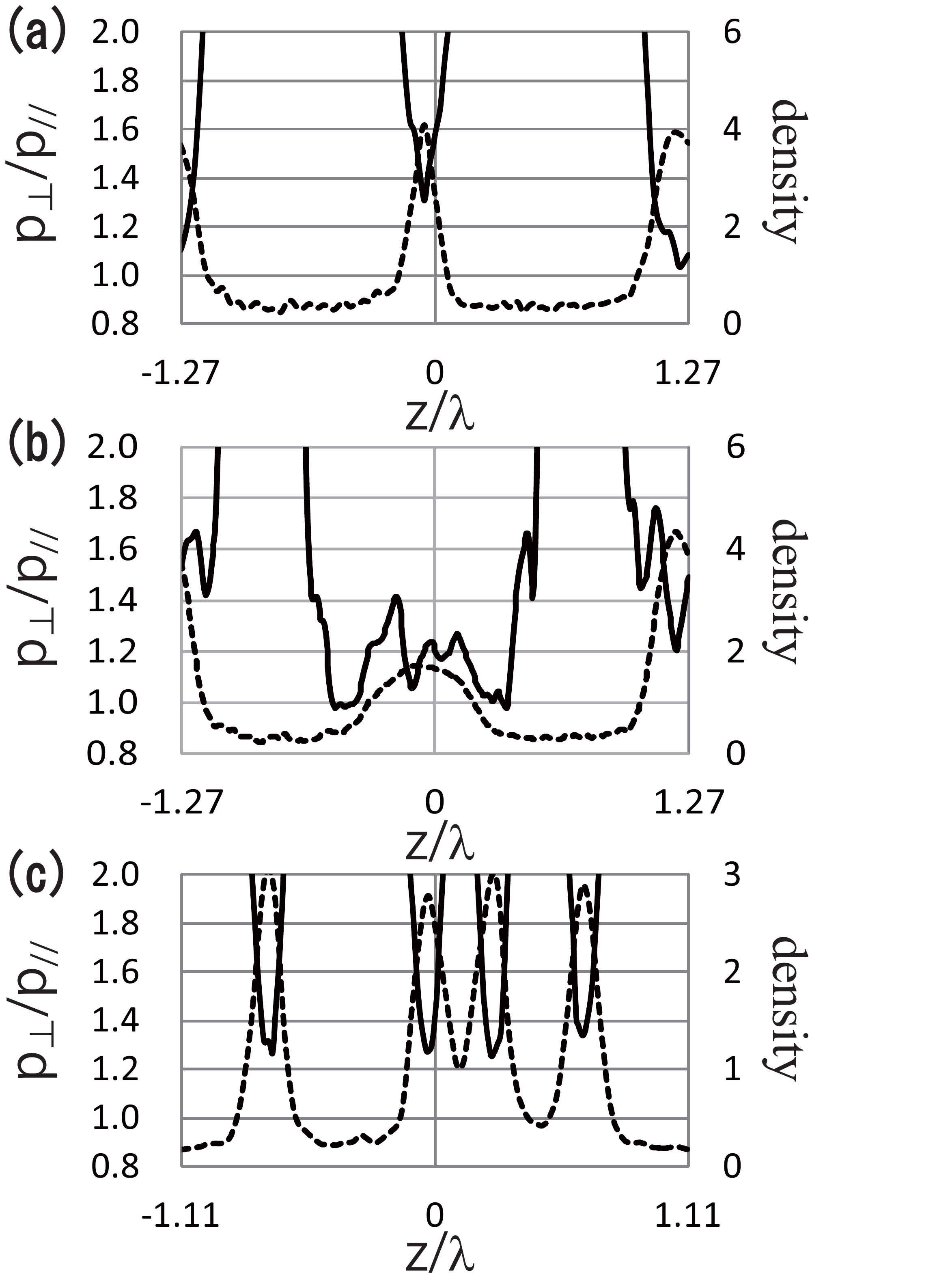}
\caption{The pressure anisotropy $p_{\perp}/p_{\para}$ integrated along the $x$ axis (solid line).  
Panels (a) and (b) are, respectively,
at $T_{\rm orbit}=5.39$ and $5.48$ for the case of $\beta=1536$.  Panel (c) is at $T_{\rm orbit}=5.16$ for the case of
$\beta=96$.  The plasma density profile (dashed line) is plotted as a reference.
\label{aniso_vs_den}}
\end{figure}

During the formation of the thin current sheet, we observed the amplified magnetic field and 
the strong pressure anisotropy outside the channel flows/the current sheet, 
and the peak value of $p_{\perp}/p_{\para}$ reached $18.7$ at $T_{\rm orbit}=5.39$.  
A pileup of the poloidal magnetic field just outside the current sheet,
as indicated by the reddish color region, can be also observed, not only outside the current sheet but also inside the current sheet. 
We can observe the pressure anisotropy of $p_{\perp} > p_{\para}$. 
The pressure anisotropy is particularly important for the onset of reconnection 
\citep[e.g.][]{Chen84,Hoshino87}.
The averaged 
pressure anisotropy and density integrated along the $x$ direction, namely,
$\int p_{\perp}(x,z) dx/\int p_{\para}(x,z) dx$ and $\int \rho(x,z) dx/(L_x \lambda)$, is shown in Figure \ref{aniso_vs_den}.
Panels (a) and (b) are at $T_{\rm orbit}=5.39$ and at $T_{\rm orbit}=5.48$, respectively.
In panel (a) we can clearly see $p_{\perp}/p_{\para} > 1.3$ for the central current sheet 
just before the onset of reconnection, and $p_{\perp}/p_{\para} \sim 1$ for the
upper current sheet.  At $T_{\rm orbit}=5.48$, after the break of the central current sheet
but before the onset of reconnection for the upper current sheet, we can find that
$p_{\perp}/p_{\para} > 1.2 \sim 1.4$, while the pressure anisotropy is 
significantly reduced for the central current sheet.
Panel (c) is the result for the lower plasma $\beta=96$.
As in the case of the evolution of $\beta=96$  (to be discussed later in detail), 
the same property of $p_{\perp}/p_{\para} > 1$ can be observed in the channel flow.

It is known that the linear growth rate of the collisionless tearing mode coupled with
the mirror mode can be given by
\begin{equation}
  \frac{{\rm Im}(\omega)}{k v_{th}} \simeq \left( \frac{p_{\perp}}{p_{\para}}-1 \right) +
  \left( \frac{r_g}{\delta} \right)^{3/2} \left( \frac{1 - k^2 \delta^2}{k \delta} \right),
\end{equation}
where $\delta$, $r_g$ and $v_{th}$  are the thickness of the current sheet, the gyro-radius,
and the thermal velocity, respectively \citep{Chen84}.  We have neglected some coefficients of the
order of unity for simplicity.  The first term on the right-hand side shows
the effect of pressure anisotropy, whereas the second term represents the standard collisionless 
tearing mode, which becomes unstable for the long-wavelength mode of $k \delta < 1$.  
The gyro-radius $r_g$ is, in general, smaller than the thickness of the current sheet $\delta$, 
then the second term is smaller than the order of unity.  Therefore, we can easily find that
the growth rate is strongly enhanced if the current sheet has a finite pressure anisotropy
of $(p_{\perp}/p_{\para} -1) > O((r_g/\delta)^{3/2})$.  In our simulation,
we found a strong pressure anisotropy of $p_{\perp}/p_{\para} \sim 1.3$, and this value
is enough to enhance dramatically the growth rate of reconnection.
We think that the rapid onset of reconnection observed during the nonlinear evolution is 
caused by the formation of pressure anisotropy during the 
stretching magnetic field of MRI.

After the onset of magnetic reconnection, the pressure anisotropy and the total magnetic field 
inside the magnetic islands decrease, while the magnetic field outside 
the islands is still increasing.  The MRI process is still very active at this time stage
through the release of magnetic field energy and pressure isotropization 
inside the islands. The reduction in pressure anisotropy $p_{\perp}/p_{\para}$ 
can be understood with the same argument based on the double adiabatic theory
described by Eqs.(\ref{CGL1}) and (\ref{CGL2}).  
Since the reconnecting magnetic field leads to magnetic field dissipation 
in the high-density region of the current sheet, the collisionless magnetic 
reconnection has a tendency to form pressure anisotropy 
with $p_{\para}/p_{\perp}>1$ in the reconnecting current sheet \citep[e.g][]{Higashi12}. 
In fact, we observe 
$p_{\perp}/p_{\para} \sim 0.3-0.4$ in the region surrounded by magnetic islands 
at $T_{\rm orbit}=6.05$.

\begin{figure}
\includegraphics[scale=0.65]{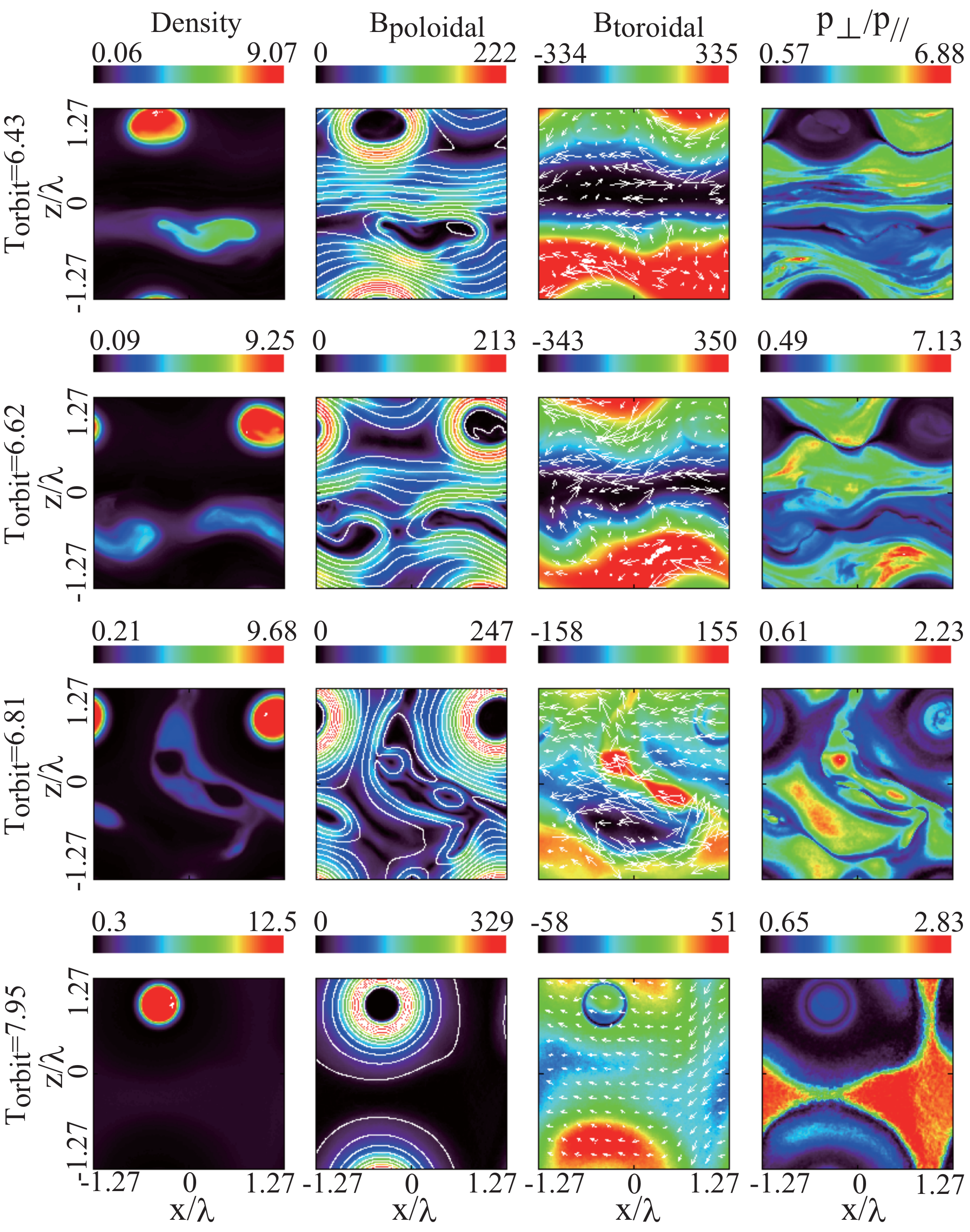}
\caption{Late evolution for $T_{\rm orbit}=6.43 \sim 6.81$ after the onset of magnetic reconnection.  
The format is the same as that of Figures \ref{fig1} and \ref{fig2}.  The magnetic islands/current sheets are subject to 
both the stretching and shrinking motions.  After that, one of the islands is deformed, and the plasma is 
spread over the entire domain.
\label{fig3}}
\end{figure}

After the break in the current sheet and the formation of magnetic islands, 
the latter islands are still subject to further MRI time evolution.  The magnetic
island situated in the center, which corresponds to the central current sheet in Figure \ref{fig2},
is stretched in the radial direction of $x$.
This kind of evolution also can be seen in the late evolution as shown in Figure \ref{fig3}.
The island situated in the lower current sheet can be further extended in the radial 
direction and forms an elongated current sheet structure at $T_{\rm orbit}=6.62$.
This time evolution of the elongating island is basically the same as that discussed by 
\citet{Hawley91}.  The elongated current sheet becomes unstable and is reconnected
at $T_{\rm orbit} =6.62 \sim 6.81$, 
and at $T_{\rm orbit}=6.81$, the island/current sheet is strongly deformed 
and is being spread over the entire region of the simulation box.  The toroidal
magnetic field and pressure anisotropy are also reduced during this stage.
At the almost-final stage of our simulation of $T_{\rm orbit}=7.95$, 
we observed a further amplified poloidal magnetic field 
$B_{\rm poloidal}/B_0 \sim  330$, while the toroidal
magnetic field was reduced to $B_{\rm toroidal}/B_0 \sim 50$.  
We could see the well-shaped magnetic island and the spread plasma gas 
surrounding the island.  We have performed several other simulation runs with 
different simulation box sizes and various plasma parameters, 
and we obtained the result that the final stage was either two magnetic islands or
a single island surrounded by the distributed plasma gas. 

\subsection{Particle Acceleration during Magnetic Reconnection}
\begin{figure}
\includegraphics{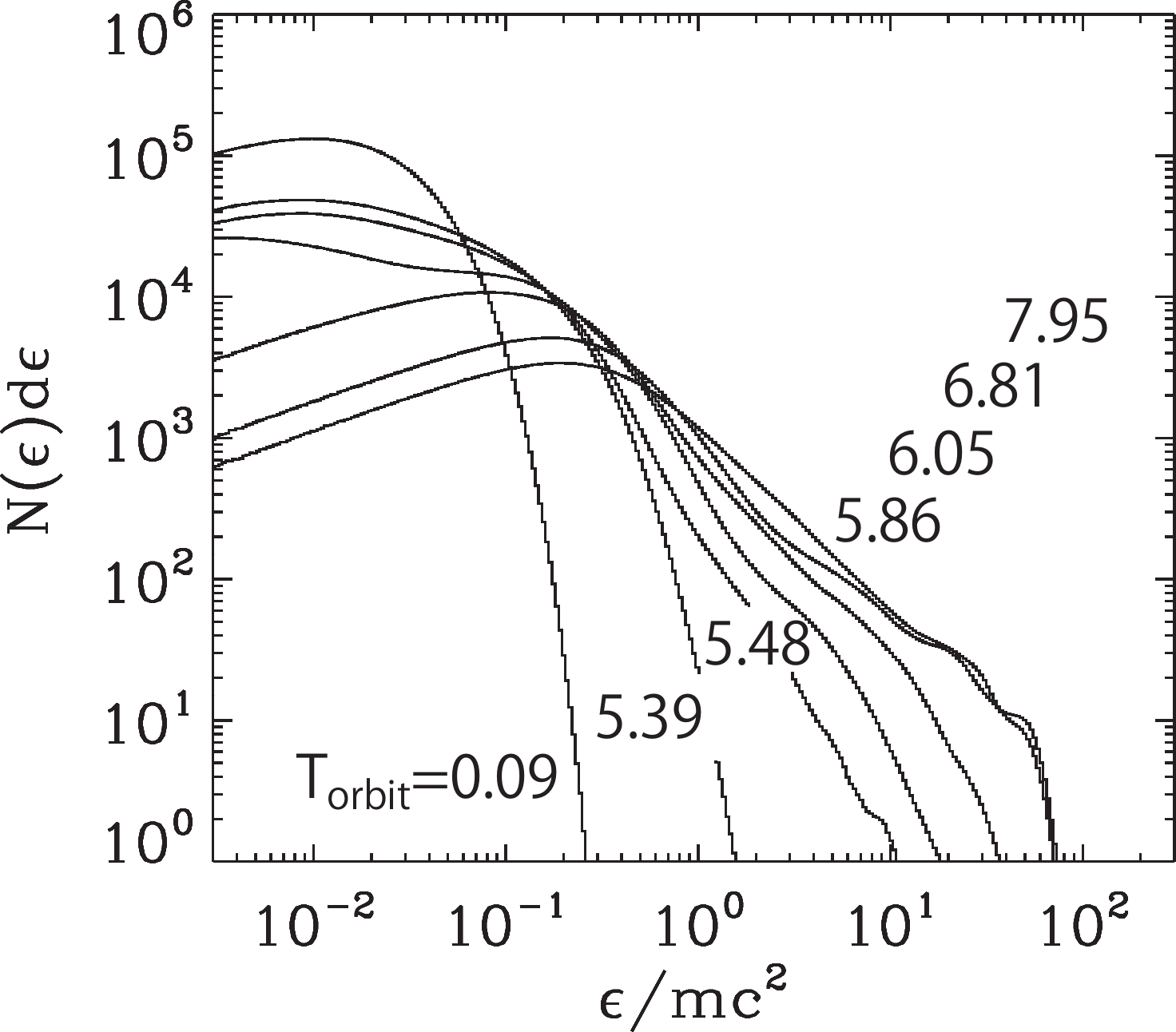}
\caption{Evolution of energy spectra for the plasma $\beta=1536$.  The initial stage at $T_{\rm orbit}=0.09$ 
shows a cold drift Maxwellian distribution function with a Keplerian motion.  
At $T_{\rm orbit}=5.39$ just before the onset of magnetic reconnection, 
the spectrum still remains a hot Maxwellian-like distribution function.  At $T_{\rm orbit}=5.48$ just after the onset of reconnection, high-energy, nonthermal particles are generated.  At $T_{\rm orbit}=7.95$ the high energy component can be approximated by a power-law function with $N(\varepsilon)d\varepsilon \propto \varepsilon^{-1}$.
\label{Espec}}
\end{figure}

Figure \ref{Espec} shows the history of the energy spectra during MRI,
which were taken at $T_{\rm orbit} = 0.09$, $5.39 \sim 6.81$, and $7.95$.
The horizontal and vertical axes are the total energy $\varepsilon/mc^2$
and the number density $N(\varepsilon)$, respectively.  
At $T_{\rm orbit}=5.39$ just before the onset of reconnection, 
the energy spectrum still had a Maxwellian-like distribution.
After the onset of reconnection at $T_{\rm orbit}=5.48$, 
nonthermal particles with a large flux were generated, 
while the thermal plasma pressure did not change much.  
During the time interval between $T_{\rm orbit}=5.86$ and $6.05$, 
further production of nonthermal particles was observed.

So far, the particle acceleration during magnetic reconnection has been discussed in numerous works.
\citet{Zenitani01} argued that the relativistic effect of the Speiser motion 
around the X-type region can form a power-law energy spectrum by drifting 
along the reconnection electric field, because the higher energetic particles
can resonate preferentially with the reconnection electric field because of the 
relativistic inertia effect.
Another important agent of acceleration is that the acceleration efficiency 
can be improved for an elongated current sheet, because the charged particle 
by the Speiser motion can remain in the acceleration region for a longer time.
As the result of both effects, the accelerated power-law energy spectrum can 
become harder \citep{HoshinoSSR12}.
The other important effect of the reconnection acceleration is the pressure anisotropy.
As discussed in the previous subsection, the pressure anisotropy of $p_{\perp}>p_{\para}$
enhances the growth rate of reconnection by coupling to the mirror mode, 
and it seems likely that the stronger induced electric field significantly contributes to 
the nonthermal particle production.

During the repeated process of disruption and formation of the current sheet,
the plasma gas could be heated and the nonthermal particles could be accelerated, 
and the energy spectrum became harder and harder.
Additionally, the nonthermal energy density increased compared with the thermal energy density.
In the almost-final stage of this simulation run at $T_{\rm orbit}=7.95$, 
the energy spectrum could be approximated by $N(\varepsilon)d\varepsilon \propto \varepsilon^{-1}$.  
The maximum energy was reached at $\varepsilon/mc^2 \sim 10^2$, for which the gyro-radius of
the highest energetic particle was almost equal to the size of the simulation box.
We will discuss this point later by compiling other simulation results.

\section{Plasma $\beta$ Dependence}
So far, we have discussed the case with $\beta=1536$ and have found that 
the pressure anisotropy plays a key role in the nonlinear evolution of MRI.
In this subsection, we study the plasma $\beta$ dependence by performing other simulations 
with $\beta=96$, $384$, and $6144$; i.e., RUNs A, B, and D as listed in Table 1.

\subsection{MRI Evolution for $\beta=96$}
\begin{figure}
\includegraphics[scale=0.65]{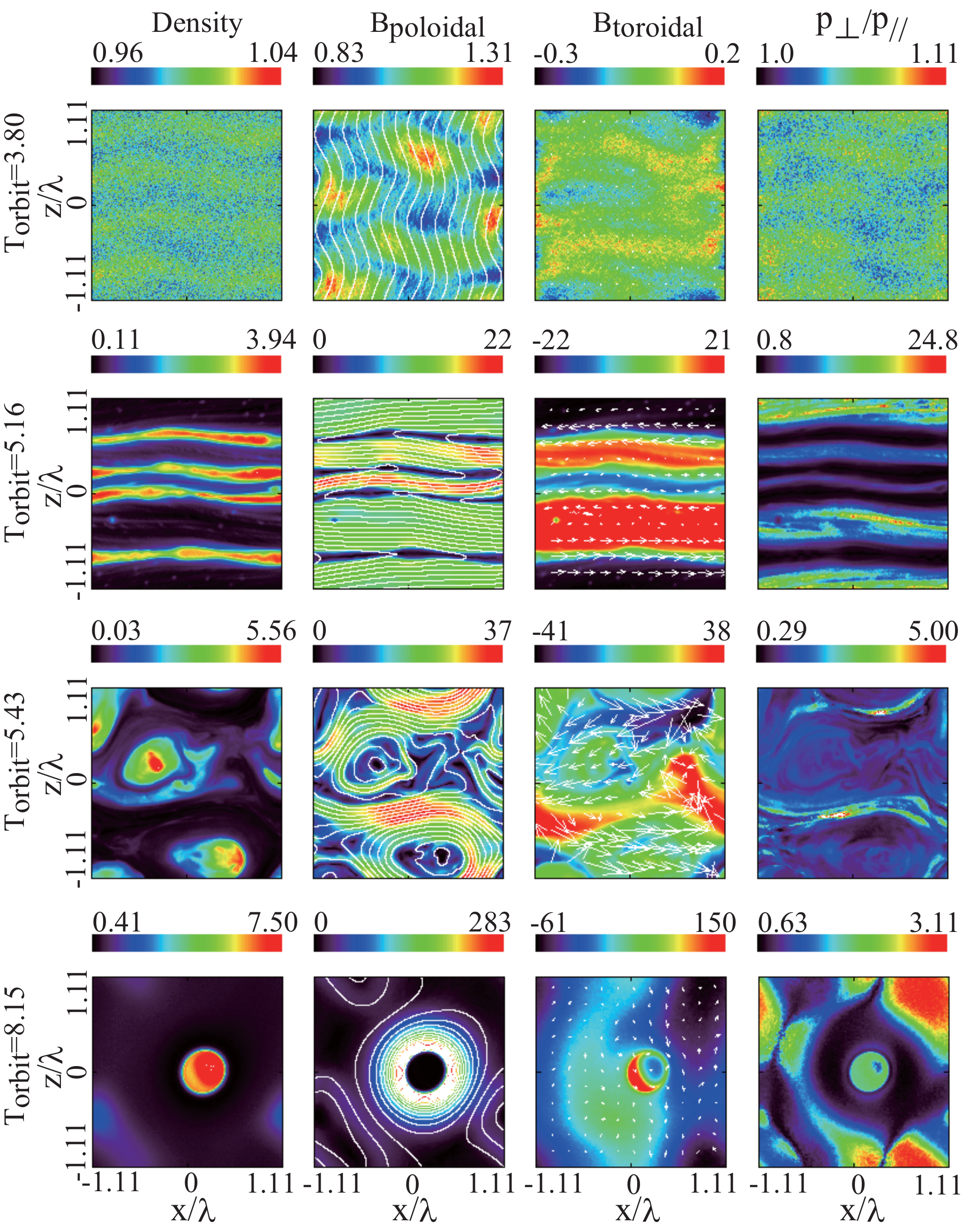}
\caption{Time evolution of MRI for plasma $\beta=96$.  The format is the same as Figure \ref{fig1}.
At the linear growth stage at $T_{\rm orbit}=3.80$, both the MRI mode propagating parallel to the rotation axis $z$ and the obliquely propagating mirror mode can be clearly seen in the poloidal magnetic field.  However, the growth of the mirror mode is relatively weak in the low plasma $\beta$ regime, and two channel flows (i.e., four current sheets) are formed at $T_{\rm orbit}=5.16$. The evolution of the late nonlinear stage after the onset of magnetic reconnection is basically the same as the higher plasma $\beta$ cases.
\label{b=96_evolution}}
\end{figure}

Figure \ref{b=96_evolution} shows the time evolution for the plasma $\beta=96$  
in the same format as Figure \ref{fig1}.
At the linear growth stage of $T_{\rm orbit}=3.80$, both the MRI mode propagating parallel 
to the rotation axis $z$ and the obliquely propagating mirror mode can be clearly seen 
in the poloidal magnetic field.  
However, the growth of the mirror mode was relatively weak during the nonlinear evolution
in the low plasma $\beta$ regime, 
and two channel flows (i.e., four current sheets) formed at $T_{\rm orbit}=5.07$.  
As the size of the simulation box was $2.28 \times (2 \pi V_A/\Omega_0)$, 
the observed unstable mode number $m=2$ was roughly consistent with the linear theory 
with isotopic pressure.   In contrast to the case with the high plasma
$\beta=1536$ in Figure \ref{fig1}, we did not observe a shift in wavelength toward 
the longer wavelength mode during the linear growth phase.

\begin{figure}
\includegraphics[scale=0.65]{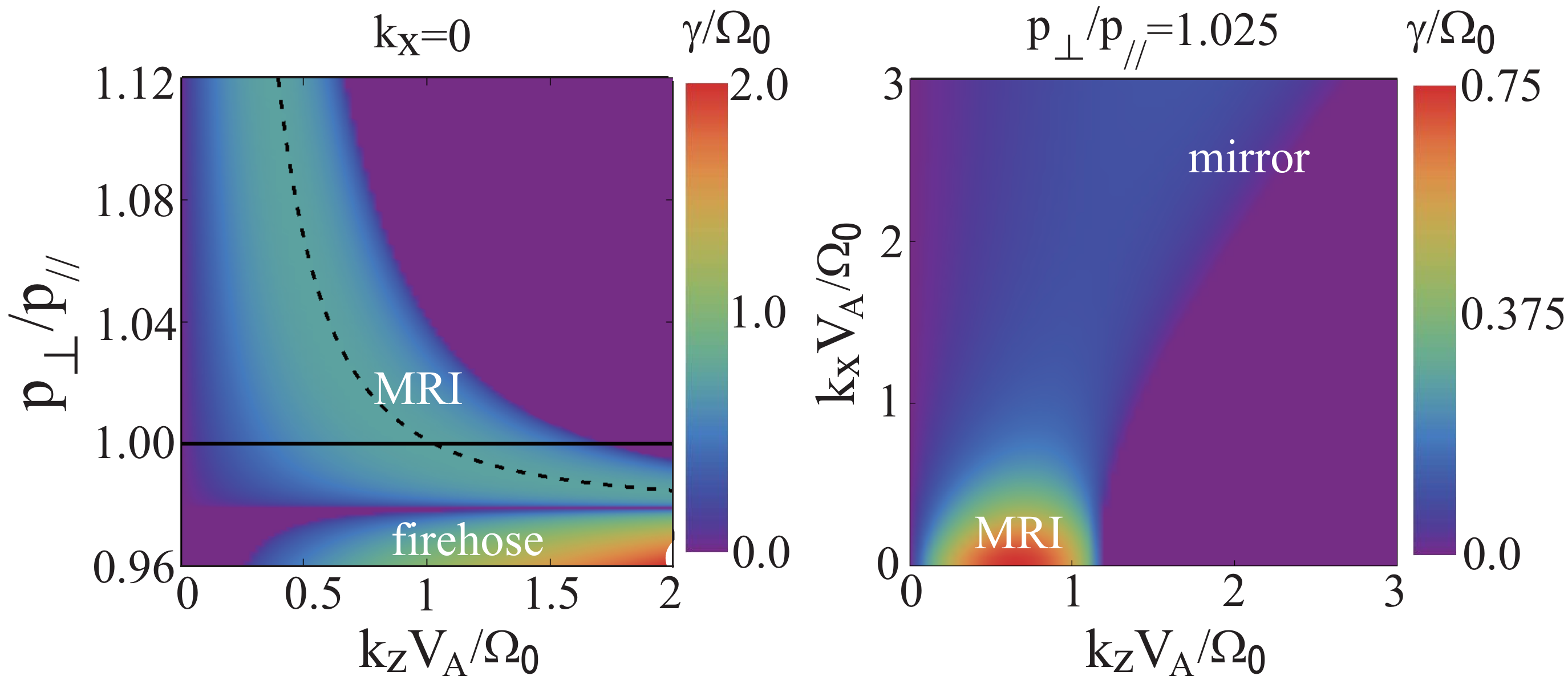}
\caption{Normalized linear growth rate $\gamma/\Omega_0$ for MRI, mirror and fire-hose modes for the plasma $\beta=96$.  
The format is same as that of Figure \ref{b=1536_theory}.  The behavior of the linear unstable modes is basically 
the same as the high plasma $\beta$ case, but the effect of the pressure anisotropy is weaker.
\label{b=96_theory}}
\end{figure}

In fact, we could confirm the weak effect of pressure anisotropy for the low
plasma $\beta$ case in the linear analysis of Figure \ref{b=96_theory}.  
The format is the same as Figure \ref{b=1536_theory}.  The same topology of the vertical 
magnetic field relative to the accretion disk was assumed.  The qualitative behavior of the instability
was the same as in the case of plasma $\beta=1536$ in Figure \ref{b=1536_theory},
but the growth rate became less sensitive to pressure anisotropy.
From the left-hand panel showing the relationship between the parallel wave number $k_z$ and
pressure anisotropy $p_{\perp}/p_{\para}$, the shift in wavelength because of
pressure anisotropy appeared for a relatively larger anisotropy, as suggested by
the linear theory of $(k_z V_A)^2 (1-(\beta_{\para}-\beta_{\perp})/2)=(15/16)\Omega_0^2$.  
To get fire-hose instability, a larger pressure anisotropy 
with $p_{\perp}/p_{\para} < 0.98$ is required as well.   The right-hand panel shows 
the coupling of the mirror mode and MRI for $p_{\perp}/p_{\para}=1.025$.
We found that the linear growth rates for the obliquely propagating mirror mode were small
compared with the high plasma $\beta=1536$ case.  This weak coupling to the mirror mode 
was consistent with the simulation result.

Let us look again at the simulation result in Figure \ref{b=96_evolution}.
The snapshot at $T_{\rm orbit}=5.43$ shows the formation of magnetic islands, and
the interaction of several magnetic islands that were formed from the four
current sheets could be seen at $T_{\rm orbit}=5.16$.  
However, the dynamic evolution of the nonlinear stage was basically the same as for
the high plasma $\beta$ case, and the almost-final stage at $T_{\rm orbit}=8.15$
had two magnetic islands.  
One of the islands contained a strong magnetic field and high plasma density, 
whereas the other had a weak magnetic field and low plasma density.
These two magnetic islands had opposite polarities of the electric current $J_y$,
and further magnetic reconnection could not occur for the two islands.

Let us discuss the pressure anisotropy in the current sheet before the onset of reconnection 
for the low plasma $\beta=96$.  The averaged pressure anisotropy at $T_{\rm orbit}=5.16$
is shown in Panel (c) of Figure \ref{aniso_vs_den}.  We observe $p_{\perp}/p_{\para} > 1.3$,
a value that is almost the same as the higher plasma $\beta=1536$ case.   We also checked
the pressure anisotropy inside the current sheets for two other runs, B and D 
with $\beta=384$ and $6144$, and found the same behavior of $p_{\perp}/p_{\para} \sim 1.3$ 
before the onset of reconnection.  We conclude that the pressure anisotropy produced 
during MRI played a significant role in the onset and rapid growth of 
reconnection.

\subsection{Total Magnetic Field, $\alpha$ Parameter, and Energy Spectra}
Let us study the amplification of the magnetic field and $\alpha$ parameter during MRI.  
So far, many MHD simulation studies have reported that the magnetic field is 
amplified during the exponential growth of the MRI 
within several rotation periods \citep[e.g.][]{BalHaw98,Sano04,Sharma06} 
and that the plasma $\beta = 8 \pi p_0/B^2$ in the saturation stage is of the 
order of $0.01 \sim 1$, where $p_0$ is defined at the initial gas pressure 
\citep[e.g][]{Hawley95}.  
We note that in the open shearing box simulation, the gas pressure continued to increase throughout the evolution 
by gaining gravitational potential energy.

\begin{figure}
\includegraphics[scale=0.5]{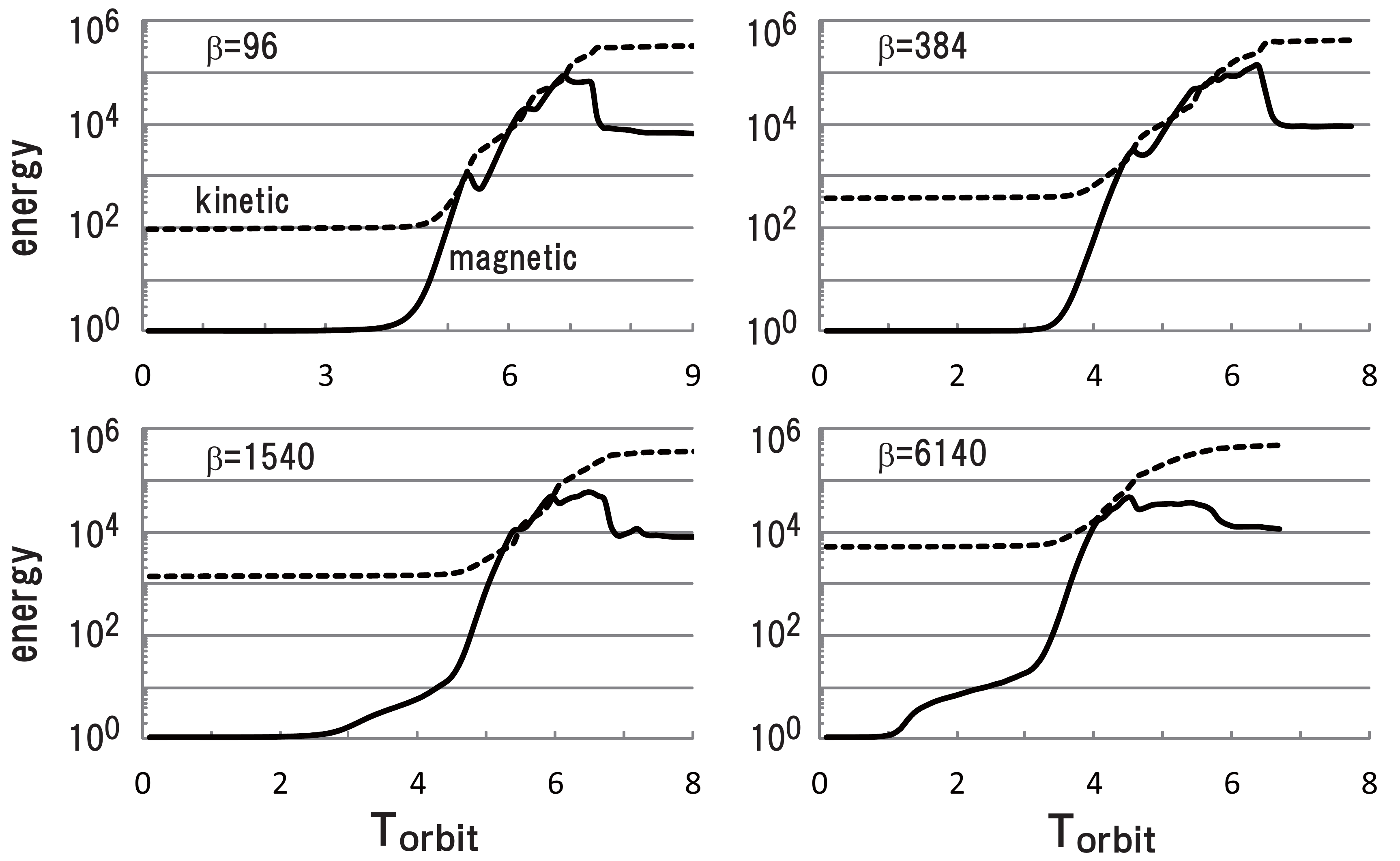}
\caption{Time history of total kinetic energy (dashed line) and magnetic field energy (solid line) 
for four different plasma $\beta$ cases with $\beta=96$, $384$, $1536$ and $6144$.  
The energies are normalized by the initial magnetic field energy.
\label{beta_dependence}}
\end{figure}

Figure \ref{beta_dependence} shows the time history of the total magnetic field energy 
and the total plasma pressure/kinetic energy integrated in the entire simulation domain.
The horizontal and vertical axes denote the orbital time and the normalized energies 
of magnetic field (solid line) and plasma pressure/kinetic energy (dashed line), respectively.
The rapid increase in magnetic field energy and plasma pressure occurs almost with the same timing 
for the lower plasma $\beta=96$ and $384$, 
but for the higher plasma $\beta=1536$ and $6144$, 
the gradual increase in magnetic field energy appears earlier than 
the rapid increase in plasma pressure.  The stages of the gradual increase in 
magnetic field correspond to the formation of the obliquely propagating mirror mode.
The timing of the rapid growth of both the magnetic field energy and plasma pressures 
corresponds to the beginning of the formation of the channel flow for all plasma $\beta$.

After the onset of the energy increase, energy equipartition can be realized in
the nonlinear phase between the magnetic and
the kinetic energies
.   However, as discussed in the previous sections,
after the onset of magnetic reconnection and the subsequent repeated processes of  
formation and deformation/disruption of the current sheet, the magnetic field energy is quickly 
released, and the plasmas are gradually heated.
In the end of the simulation, we found that 
the plasma $\beta=8 \pi p_0/B^2$ was of the order of $0.01 \sim 1$, which coincides with
earlier MHD simulation experiments. 
The plasma $\beta=8 \pi p/B^2$ defined by using the instantaneous pressure becomes 
several 10 regardless to the amplitude of the initial magnetic field.  
It is interesting to note that the amplification of 
the total magnetic field was about $10^4$ times longer than the initial state 
for all our simulations, but this result might depend on
the simulation box size, because the size of the final magnetic island probably depends on the box size.

\begin{figure}
\includegraphics{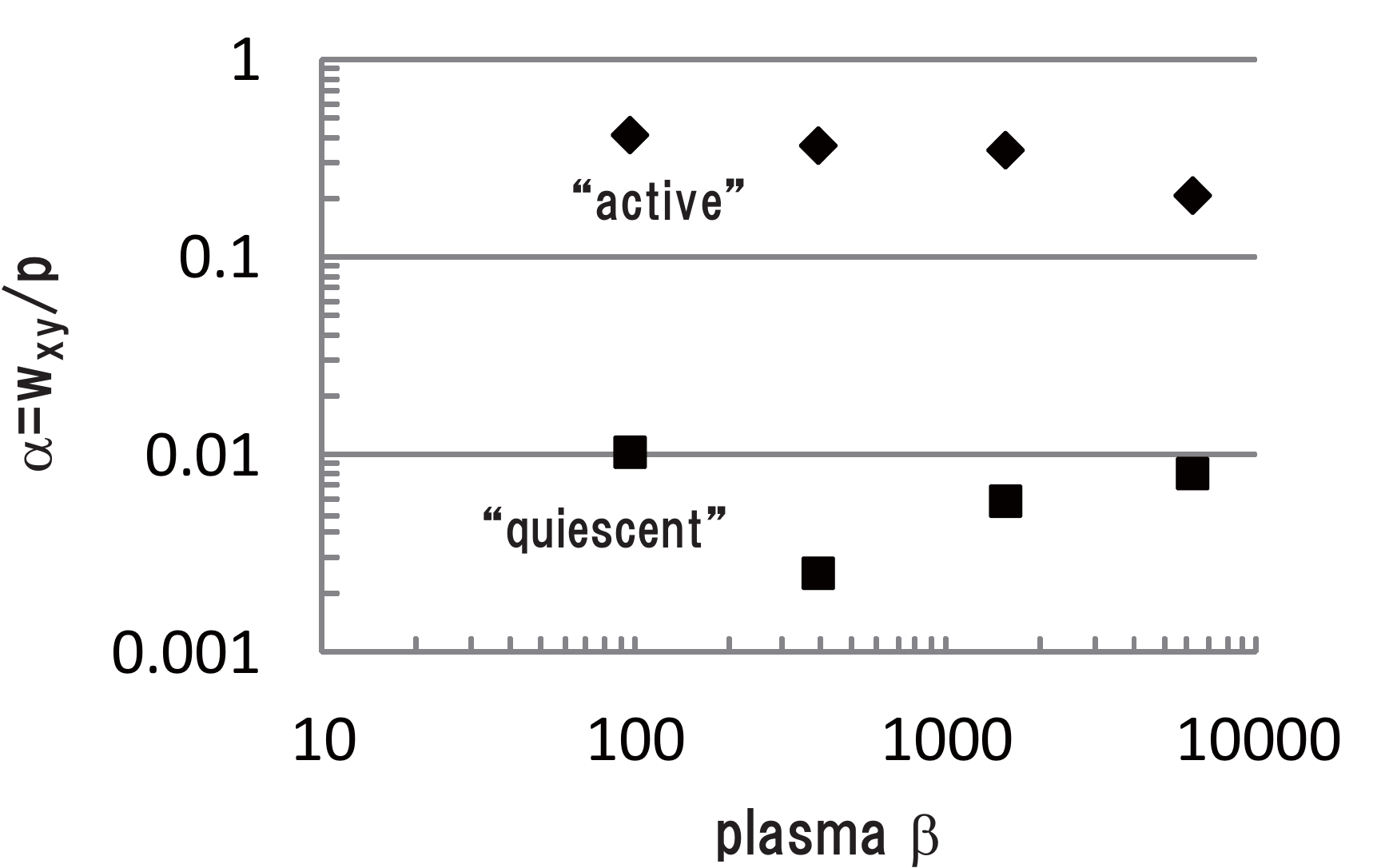}
\caption{The $\alpha$ parameter measured during the ``active'' reconnection phase 
and at the ``quiescent'' final phase for four different plasmas with 
$\beta=96$, $384$, $1536$ and $6144$.
\label{alpha}}
\end{figure}

Let us study the angular momentum transport in our two-dimensional simulations by 
numerically measuring the stress tensor given by
\begin{equation}
  w_{xy}=\rho v_x (v_y + q \Omega_0 x)-\frac{B_x B_y}{4 \pi} + \frac{(p_{\para}-p_{\perp})}{B^2}B_x B_y.
\end{equation}
Each of the terms on the right-hand side represents the Reynolds ($w_{R}$), Maxwell ($w_{M}$) 
and anisotropy ($w_{P}$) stresses, respectively \citep{Sharma06}.  
As the above stress tensor is related to the energy dissipation rate in the system, 
the so-called $\alpha$ parameter in the standard accretion disk model \citep{Shakura73}
can be defined by
\begin{equation}
  \alpha = \frac{w_{xy}}{p}=\frac{w_{R}+w_{M}+w_{A}}{p},
\end{equation}
where $p$ is the plasma pressure. 
We measured the $\alpha$ parameters during the active reconnection stage and during the 
almost-final stage with a relatively quiet phase.  For instance, the ``active'' reconnection stage
corresponds to the time interval of $T_{\rm orbit}=6 \sim 7.5$, whereas the almost-final, ``quiescent'' stage
is $T_{\rm orbit}=7.5 \sim 9$ for the case with $\beta=96$.  Namely, we defined the active reconnection phase 
by the time interval during which the magnetic energy is greater than that in the almost-final stage. 
We found that $w_{M} > w_{R} > w_{P}$ during the active phase, and
$w_{M} > w_{p} \ge w_{R}$ during the almost-final phase in our simulation results.
Figure \ref{alpha} shows the measured $\alpha$ parameters for four different plasma $\beta$.
The diamond and square symbols stand for the ``active'' phase and the 
``quiescent'' final phase, respectively.  
We found that $\alpha \sim 10^{-1} \sim 10^0$ during the active phase, while $\alpha$ was about
$10^{-3} \sim 10^{-2}$ during the almost-final stage.  Roughly speaking, these values seem to support
the standard $\alpha$ disk model \citep{Shakura73}.

\begin{figure}
\includegraphics{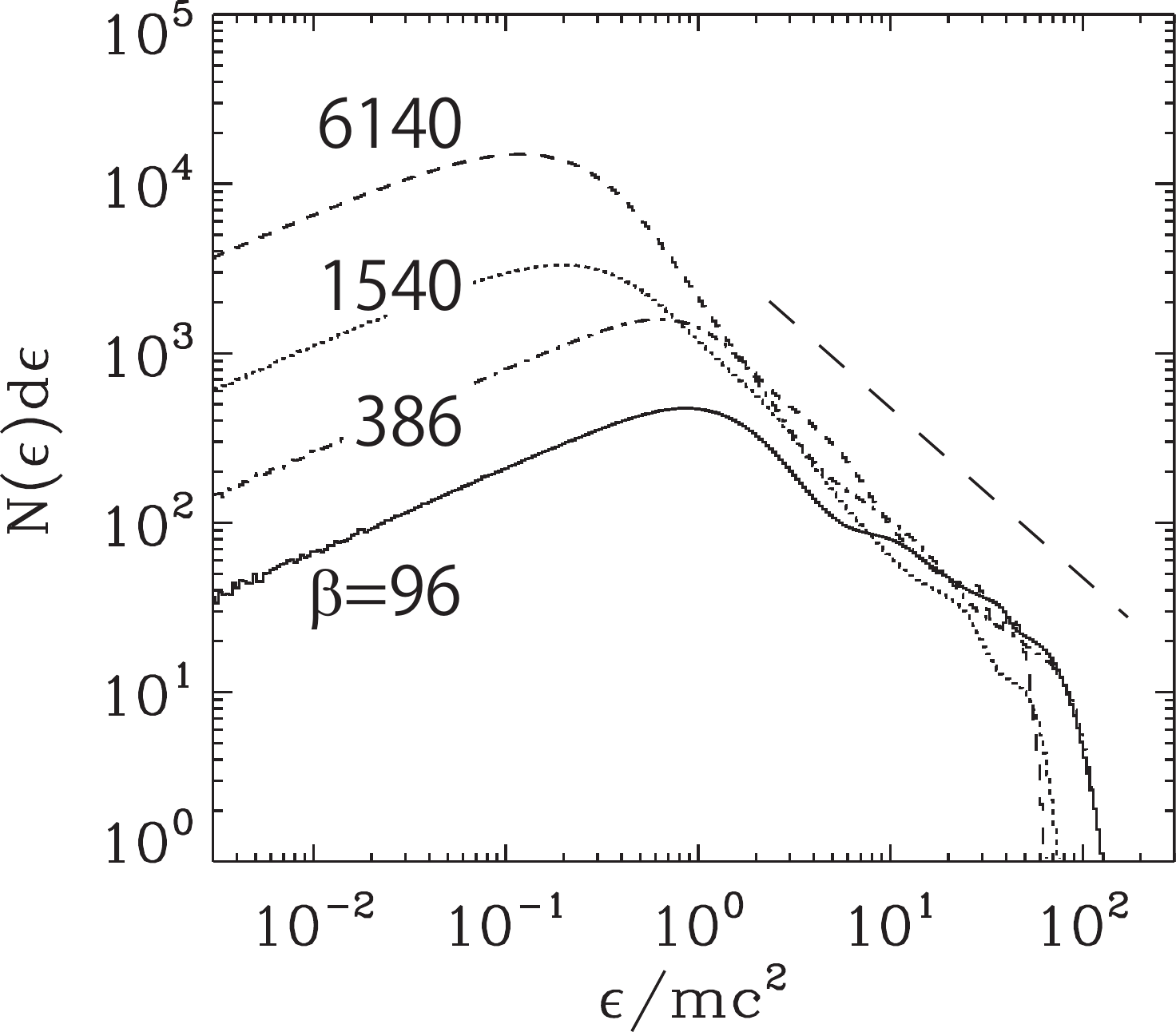}
\caption{Energy spectra of the almost-final phase 
for four different plasma $\beta=96$, $384$, $1536$ and $6144$.  
The format is the same as that of Figure \ref{Espec}.  
All high-energy tails can be approximated by a power-law function with 
$N(\varepsilon) d\varepsilon \propto \varepsilon^{-1}$ shown in the long dashed line.
\label{Espec_all}}
\end{figure}

Finally, we examined the energy spectrum and the maximum attainable energy.
The energy spectra obtained from the almost-final
stages of four plasma $\beta$ RUNs with $\beta=96$, $384$, $1536$ and $6144$ are shown in Figure \ref{Espec_all}.  
Regardless of plasma $\beta$, the well-developed nonthermal spectra
can be formed during MRI, and the high-energy tail can be approximated by 
a power-law spectrum of $N(\varepsilon)d\varepsilon \propto \varepsilon^{-1}$.  
The maximum attainable energies are probably limited by the gyro-radius of the simulation
box size, and the maximum energy $\varepsilon_{\rm max}$ can be expressed by
\begin{equation}
   \frac{\varepsilon_{\rm max}}{mc^2} \sim \frac{e B_{f}}{mc^2} (L_x \lambda) 
   \sim \left(\frac{V_A}{c}\right) \left(\frac{\Omega_c}{\Omega_0}\right) 
        \left(\frac{B_f}{B_i}\right) \left(2 \pi L_x \right),
\end{equation}
where $B_f/B_i$ is the ratio of the magnetic field amplification, and $L_x$ is 
the normalized simulation box size normalized by $\lambda=2 \pi V_A/\Omega_0$.  
As we can see in Figure \ref{beta_dependence},
the ratio of $B_f/B_i$ is about $10^2$ regardless of plasma $\beta$.  Referring to
the plasma parameters of $V_A/c$ , $\Omega_c/\Omega_0$ and $L_x$ listed in Table \ref{tbl1}, 
we can obtain the maximum attainable energy 
$\varepsilon_{\rm max}/mc^2 \sim 174$, $150$, $100$, and $75$ 
for the corresponding $\beta=96$, $384$, $1536$, and $6144$, respectively.  
The high-energy cutoffs in Figure \ref{Espec_all} seem to be consistent with
the above simple discussion.

\section{Discussion and Summary}
We discussed the nonlinear time evolution of MRI in a collisionless accretion disk, 
where the mean-free path is longer than the typical scale size of the disk.  
We found that the collisionless MRI can produce significant amounts of nonthermal particles
during the saturation stage through magnetic reconnection.  The pressure anisotropy of 
$p_{\perp}/p_{\para} > 1$ initiated during the nonlinear evolution of MRI can
lead to a rapid onset of magnetic reconnection by coupling of the tearing and
mirror modes \citep{Chen84}, 
and its strong inductive electric field excited by the fast reconnection
can enhance the efficiency of particle acceleration.

In our simulation experiments, we used a relatively small simulation box, but 
it is supposed that many magnetic islands may form in a large accretion disk system  
and that further nonthermal acceleration may occur.  As one of the possible acceleration processes in many magnetic islands,
the stochastic ``Fermi-reconnection acceleration'' has been discussed recently by \citet{HoshinoPRL12}.  
In the standard Fermi acceleration model \citep{Fermi49}, 
charged particles gain energy stochastically during head-on 
and head--tail collisions of particles with magnetic clouds as the scattering objects,
and the increase in particle energy is known to be the second order of $V_c/c$, 
where $V_c$ and $c$ are the velocity of the random motion of the magnetic clouds and 
the speed of light, respectively.  
On the other hand, \citet{HoshinoPRL12} proposed the stochastic Fermi-reconnection
acceleration by introducing the interaction of charge particles with many magnetic islands 
instead of the magnetic clouds, and discussed that the acceleration efficiency becomes first order 
of $V_A/c$.

\begin{figure}
\includegraphics{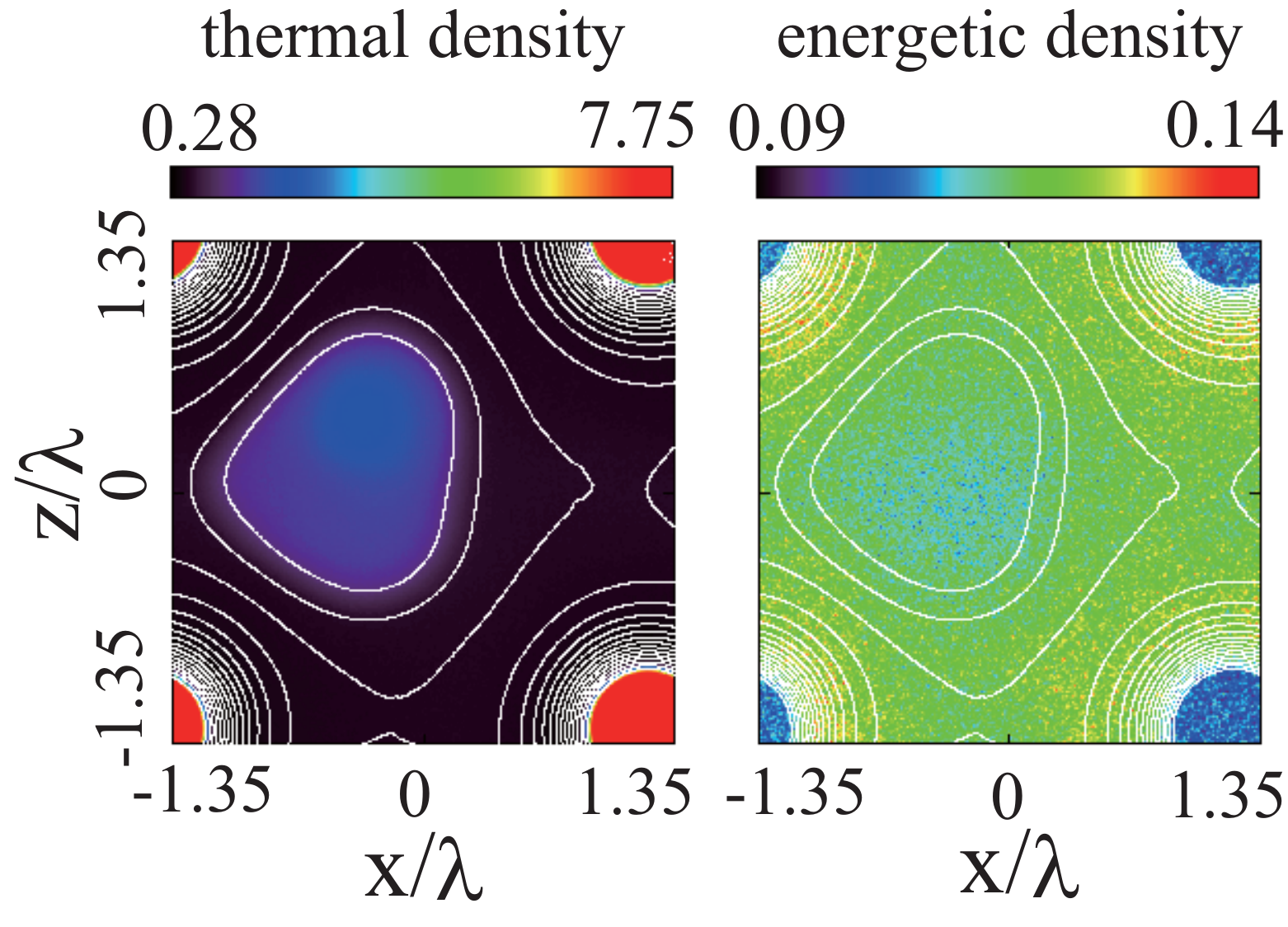}
\caption{The color contours of the densities of the thermal plasma (left) and the energetic particles (right).  
The white lines show the magnetic field lines.
\label{energetic_density}}
\end{figure}

To check the possibility of the stochastic Fermi-reconnection acceleration,
we examined the relationship between the thermal plasma density and the energetic particle density in the system.
Figure \ref{energetic_density} shows the color contours of the thermal plasma density 
and the higher energy density for RUN B with $\beta=384$ at $T_{\rm orbit}=7.26$.  
We calculate the densities by integrating the velocity distribution function with the particle
momentum according to $\int_{\varepsilon_{min}}^{\varepsilon_{max}} f(\varepsilon)d\varepsilon$.  
The left-hand and right-hand panels show the densities of ``thermal'' plasma with 
$(\varepsilon_{min},\varepsilon_{max})/mc^2=(0,\infty)$ and 
``energetic'' plasma with $(50,\infty)$, respectively.
In contrast to the thermal plasma density, we see that the ``energetic'' plasma forms a void structure 
inside the island, and the particles are distributed outside the magnetic islands.  
Not only the main magnetic island situated around the corner but also the weaker magnetic island 
located in the center shows the same tendency of preferential distribution of the particles.
This nonuniform density distribution can enhance the probability of the interaction of head-on
collisions with the reconnection outflows and leads to the efficient first-order acceleration.
Therefore, we think that the stochastic Fermi-reconnection acceleration can occur during MRI
in a large-scale system.

We showed the formation of large pressure anisotropy during MRI, but the pressure anisotropy
may be exaggerated.  In a realistic accretion disk, the cyclotron frequency $\Omega_c$ should
be much larger than the disk rotation frequency $\Omega_0$, but we used $\Omega_c/\Omega_0=10$
in this paper.  Because the mirror-mode instability may occur in the time scale of $O(10 \Omega_c^{-1})$,
the relaxation process of the pressure anisotropy during the disk rotation of $\Omega_0^{-1}$ 
should be much more effective than in our simulations.
However, the pressure anisotropy generated inside the current sheet/channel flow may not be
exaggerated.  Roughly speaking, 
the time scale of the evolution of magnetic reconnection for a thin current sheet is
of the same order of the mirror/ion-cyclotron instability.
Therefore, the growth of reconnection can compete with the mirror/ion-cyclotron instability. 
More importantly, \citet{Gary94,Gary97} reported that a finite pressure anisotropy may remain
in a collisionless plasma, and the threshold condition for the pitch-angle scattering of 
mirror/ion-cyclotron instability may be given by
$p_{\perp}/p_{\para} -1 \sim {\rm Min}(7/\beta_{\perp},0.35/\beta_{\para}^{0.42})$.
This threshold value is small, but it may significantly affect on the growth of magnetic reconnection.

We evaluated the stress tensor during MRI and obtained that the value of $\alpha$ normalized by the
heated plasma pressure was of the order of $O(10^{-3}) \sim O(10^{-1})$. However,
if we normalize the stress tensor by the initial plasma pressure $p_0$, the $\alpha$
parameter becomes of the order of $O(10^{-1}) \sim O(10^{2})$, which 
suggests a much more efficient angular momentum transport than the one discussed previously
\citep[e.g.][]{Hawley95,Sano04}.  
The reason for the large $\alpha$ value during the active reconnection phase
is probably the fact that the reconnection process can couple to the mirror mode under 
the pressure anisotropy of $p_{\perp}>p_{\para}$.  The rapid reconnection may lead to
strong fluctuation/turbulence that can contribute to the enhancement of the stress tensor.
However, as our simulation study was performed in a two-dimensional system with a
relatively small simulation box, rich nonlinear wave-coupling processes that probably
occur in a large-scale full-dimensional system seem to be limited. 
A careful examination of this issue will need to be performed in a future study.

We observed the repeated processes of current sheet formation by stretching the magnetic field 
and the deformation/disruption of a current sheet by reconnection, 
and during the repeated processes with the active reconnection, 
strong turbulence was observed.  However, after several repeated processes, 
the fluctuation levels were reducing with time, and we obtained a more or less 
``quiescent'' state, where the plasma flow fluctuations were relatively weak.  
We discuss the two-dimensional behavior of MRI in the collisionless system in this paper, 
but needless to say, the study of a three-dimensional simulation is needed.  
In general, the plasma transport processes may be different between two- and three-dimensional systems, 
and the saturation of MRI and the angular momentum transport studied in this paper would be modified
in three-dimensional MRI.  We will discuss three-dimensional evolution of MRI in another paper.

Our simulation study has several limitations, but a couple of important implications may be given 
for interpreting the observed high-energy particle emission from massive black holes such as Sgr A*
\citep[e.g.][]{Nara98,Quat03,Yuan03,Aharonian08,Chernyakova11,Kusunose12}.
Magnetic reconnection involved in a collisionless accretion disk can be regarded as 
a plausible process to generate not only high-energy particles but  also MHD turbulence, 
because a small but finite pressure anisotropy of $p_{\perp}>p_{\para}$
induced by MRI plays an important role in the onset and growth of reconnection.
By virtue of the active reconnection caused by the coupling between the tearing and mirror modes, 
a strong inductive reconnection electric field enhances the particle acceleration efficiency, 
and the angular transport may be enhanced during the active reconnection phase through 
strong turbulence as well.  The collisionless reconnection with $p_{\perp}>p_{\para}$ 
might lead to an efficient mass accretion to maintain a collisionless accretion disk.
 
\acknowledgments
This work was supported in part by JSPS Grant-in-Aid for Scientific Research (KAKENHI) 
Grant Number 25287151.  The author thanks R. Matsumoto, T. Sano, S.-I. Inutsuka, 
K. Hirabayashi and K. Shirakawa for valuable discussions.

\appendix
\section{Linear Dispersion for MRI in Pair Plasmas}
We perform a linear analysis of magnetorotational instability in pair plasmas 
and show that the basic property of MRI instability in a pair plasma is the same as 
that discussed in a one-fluid MHD system.
A set of the basic equations in a pair plasma in a local rotating disk 
with the angular velocity $\vec{\Omega}_0=(0,0,\Omega_0)$ can be given in the Cartesian frame as follows:
\begin{eqnarray}
 \frac{\partial n_\pm}{\partial t} &=& - \nabla \cdot (n_\pm \vec{v}_\pm) , \\
 \left( \frac{\partial}{\partial t}+\vec{v}_\pm \cdot \nabla \right) \vec{v}_\pm &=& -\frac{1}{n_\pm m_\pm} \nabla p_\pm + 
  \frac{e_\pm}{m_\pm} \left( \vec{E}+\frac{\vec{v}_\pm}{c} \times \vec{B} \right) - 
  2 \vec{\Omega}_0 \times \vec{v}_j + 2 q \Omega_0 x \vec{e}_x, \\
 \frac{1}{c}\frac{\partial \vec{B}}{\partial t} &=& - \nabla \times \vec{E}, \label{faraday2} \\
 \nabla \times \vec{B} &=& \frac{4 \pi}{c} \left(e_+ n_+ \vec{v}_+ -|e_-| n_- \vec{v}_- \right),  \label{ampere2} \\
 \frac{D}{Dt}\left(\frac{p_\pm}{n_\pm^\gamma} \right) &=& 0.
\end{eqnarray}
We assume that the initial magnetic field has only a $z$ component, i.e., $\vec{B_0}=(0,0,B_0)$, 
gas pressure $p_\pm$ is constant, and the initial Keplerian velocity shear has a y component, 
i.e., $\vec{v}_{0,\pm} =(0, -q \Omega_0 x,0)$, 
where $q=-\partial {\rm ln} \Omega/\partial {\rm ln} r = 3/2$, 
With this assumption, the electric field should satisfy the condition of 
$\vec{E}=-\vec{v}_{0,\pm} \times \vec{B}_0/c$.  Because of this finite electric field, however,
the initial number density of $n_\pm$ is not exactly the same, but the charge difference of 
$(n_+ - n_-)/n_+ \sim O((V_A^2/c^2)(\Omega_0/\Omega_c)) \ll 1$ is neglected in this analysis.

In the linearized equations, we also assume the charge neutrality with $\delta n_+ = \delta n_-$.
We define the sum and difference of the perturbed fluid velocities for the positron and electron as
\begin{eqnarray}
  \delta \vec{u} &\equiv& (\delta \vec{v}_+ + \delta \vec{v}_-)/2, \\
  \delta \vec{w} &\equiv& (\delta \vec{v}_+ - \delta \vec{v}_-)/2. 
\end{eqnarray}
By this definition, the sum and difference of the momentum equation yield
\begin{eqnarray}
 -i \omega \delta \vec{u} -\Omega_0 q \delta u_x \vec{e}_y &=& \delta \vec{w} \times \Omega_c \vec{e}_z 
      -2 \Omega_0 \vec{e}_z \times \delta \vec{u} - i V_s^2 \vec{k} \delta n/n ,  \label{u_equation} \\
 -i \omega \delta \vec{w} -\Omega_0 q \delta w_x \vec{e}_y &=& \delta \vec{u} \times \Omega_c \vec{e}_z 
      -2 \Omega_0 \vec{e}_z \times \delta \vec{w}  
      + \frac{e}{m} \delta \vec{E} + \Omega_c v_0 \vec{e}_y \times \delta \vec{B} / B_0, \label{w_equation}
\end{eqnarray}
and the Amp\`ere's equation (\ref{ampere2}) can be written as
\begin{equation}
  \delta \vec{w} (\Omega_c/V_A^2) = i \vec{k} \times \delta \vec{B}/B_0. \label{Ampere_equation}
\end{equation}

To eliminate $\delta \vec{E}$, we substitute Eq.(\ref{w_equation}) into Faraday's equation (\ref{faraday2})
and use Eqs.(\ref{u_equation}) and (\ref{Ampere_equation}). 
Assuming the linear perturbation form of ${\rm exp}(i \vec{k} \cdot \vec{x} - i \omega t)$ with
the wave vector $\vec{k}=(k_x,0,k_z)$, we finally obtain the dispersion relation as
\begin{eqnarray}
\left( \omega^2 (1+\varepsilon) -k_z^2 V_A^2(1-(q-2)^2 \nu) \right)
\left( (1+\varepsilon) \omega^2 \frac{\omega^2-k^2 V_S^2}{\omega^2-k_z^2 V_S^2}
   -k^2 V_A^2+\Omega_0^2 (2-q) \varepsilon \right) \nonumber \\
=  \Omega_0^2 (2-q) \left( 2 (1+\varepsilon)+ \varepsilon (2-q)  \right) \times \nonumber \\
\left( \omega^2 (1+\varepsilon)+ \frac{q}{2-q} k_z^2 V_A^2+ k_z^2 V_A^2 
 \left((q-2) + \omega^2 \frac{\omega^2-k^2 V_S^2}{\omega^2-k_z^2 V_S^2} \right)\nu \right),
\end{eqnarray}
where $\varepsilon=k^2 V_A^2/\Omega_c^2$, $\nu=\Omega_0^2/\Omega_c^2$, and $k=\sqrt{k_x^2+k_z^2}$.
In the limit of both $\varepsilon \rightarrow 0$ and $\nu \rightarrow 0$, the above dispersion relation
can be simplified as
\begin{equation}
\left( \omega^2 -k_z^2 V_A^2 \right)
\left( \omega^2 \frac{\omega^2-k^2 V_S^2}{\omega^2-k_z^2 V_S^2}- k^2 V_A^2 \right)
=  2 \Omega_0^2 (2-q) \left( \omega^2 + \frac{q}{2-q} k_z^2 V_A^2 \right).
\end{equation}
The first term on the left-hand side shows the shear Alfv\'en mode, and the second term represents
the slow and fast modes, whereas the right-hand side is the coupling term caused by the differential rotation.
This dispersion equation is exactly the same as that obtained in the MHD system.  Therefore, the linear MRI behavior
in a pair plasma should be same as that in a one-fluid MHD equation.

\section{An Equilibrium Solution for the Vertical Magnetic Field Geometry}
For our kinetic MRI study in the meridional plane, the initial magnetic field $B_z$ is 
orthogonal to the Kepler velocity $v_y$, and then a finite motional electric field 
$E_x$ appears, which varies in the radial direction.  Therefore, a finite electric charge 
density $\rho_c$ may appear as well.  The order of magnitude of the charge density is
\begin{equation}
  \frac{\rho_c}{e n_\pm} \sim \frac{1}{c e n_\pm} \nabla \cdot (v_y \times B_0)
  \sim \frac{V_A^2}{c^2} \frac{\Omega_0}{\Omega_c}, 
\end{equation}
where $V_A^2 = B^2/4 \pi n m$ and $\Omega_c = e B/ m c$, and 
we can easily find that this charge separation is very small in the low-frequency MHD regime.

To obtain a better initial condition including charge separation, however,
we solve the coupled Amp\`ere's and Poisson's equations as follows:
\begin{eqnarray}
  4 \pi \rho_c(x) 
     &=& \nabla \cdot \left( \vec{E}(x)-\frac{\vec{v}_0}{c} \times \vec{B}_z \right) 
      = -\frac{1}{c} \nabla \cdot \left( (v_y(x) + v_0(x)) \vec{e}_y \times B_z(x) \vec{e}_z \right), \\
  v_y(x) \rho_c(x) &=& \vec{J}(x) = \frac{c}{4 \pi} \nabla \times (B_z(x) \vec{e}_z), \label{ampere}
\end{eqnarray}
where $v_y(x)=-q \Omega_0 x$ with $q=3/2$ for a Keplerian disk, 
and $\vec{v}_0(x)=\Omega_0 (x+r_0)$ is the term arising from the noninertial frame.  
By eliminating the charge density, we obtain
\begin{equation}
 \frac{1}{B_z(x)} \frac{\partial B_z(x)}{\partial x} = \frac{\alpha x}{1+\beta x -\alpha x^2},
\end{equation}
where $\alpha = \Omega_0^2 q(q-1)/c^2$ and $\beta=\Omega_0^2 q r_0/c^2$.
This differential equation can be solved analytically, but by assuming $\alpha \ll \beta \ll 1$, we have the solution as
\begin{equation}
  B_z(x)=\frac{B_0}{\sqrt{1+\beta x -\alpha x^2}} \sqrt{\frac{| \alpha x |}{| \beta -\alpha x |}}.
\end{equation}
Furthermore, if we neglect the noninertial term of $v_0(x)$, the solution becomes
\begin{equation}
  B_z(x)=B_0/\sqrt{1- \alpha' x^2}, \label{b_solution}
\end{equation}
where $\alpha'=(q \Omega_0/c)^2$.  We basically used this form as the initial condition in our simulations.
Note that the simulation domain is much less than the light cylinder $c/\Omega_0$, 
and then $\alpha' x^2 \ll 1$.

Next, let us obtain the corresponding distribution of the finite charge density $\rho_c(x)$. 
We have assumed the same number density between positrons and electrons, i.e.,
$n_+(x)=n_-(x)=n(x)$, but a different charge density is adopted between electrons and positrons;
i.e., $e_+ \ne |e_-|$ by keeping $m_+/m_-= e_+/|e_-|$ in our simulation.  
By using Eqs.(\ref{ampere}) and (\ref{b_solution}),  
the corresponding charge density correction can be obtained by
\begin{equation}
  \frac{\rho_c(x)}{e_+ n_0}=\left( \frac{e_+-|e_-|}{e_+} \right) \frac{n(x)}{n_0}
      =\frac{B_0 q \Omega_0}{4 \pi e_+ n_0 c} \frac{1}{(1-\alpha' x^2)^{3/2}},
\end{equation}
and we adopt the solutions of
\begin{equation}
  \frac{e_+-|e_-|}{e_+} =\frac{B_0 q \Omega_0}{4 \pi e_+ n_0 c},
\end{equation}
and
\begin{equation}
  \frac{n(x)}{n_0} =\frac{1}{(1-\alpha' x^2)^{3/2}}.
\end{equation}

In addition to these corrections of magnetic field and charge density, the plasma gas pressure
is adjusted to satisfy the pressure balance of 
\begin{equation}
  \frac{B_z^2(x)}{8 \pi} + n(x) \left( T_+(x) + T_-(x) \right) = \rm{const}.
\end{equation}
In our equilibrium solution, plasma pressure is also a function of $x$.  This solution is
an equilibrium solution in the level of fluid approximation, 
but it would not be necessarily a Vlasov equilibrium.
We used $V_A/c \sim 10^{-2} - 10^{-3}$ and $\Omega_0/\Omega_c=10^{-1}$ 
in this paper, so that these correction terms are of the order of $10^{-5}$.
Therefore, no significant difference may appear for the nonlinear evolution 
with or without the correction terms.
\section{Simulation Code}
We briefly describe our simulation code, which was slightly modified from the STARFIELD code \citep{Hoshino87,Hoshino92}.
The velocities and positions of particles were integrated in time by using 
the standard Buneman--Boris methods; namely,
\begin{eqnarray}
 \frac{\vec{p}^{n+1/2}-\vec{p}^{n-1/2}}{\Delta t} &=& 
    e \left(\vec{E}^{n}_{\rm mod}+\frac{\vec{v}^{n+1/2}+\vec{v}^{n-1/2}}{2c} \times \vec{B}^{n}_{\rm mod} \right) \\
 \frac{\vec{x}^{n+1}-\vec{x}^{n}}{\Delta t} &=& \vec{v}^{n+1/2},
\end{eqnarray}
where $\vec{p}=\gamma m \vec{v}$, and a suffix $n$ shows a time step.  The modified $E_{\rm mod}$ and $B_{\rm mod}$ terms 
by the Coriolis and tidal forces are given by
\begin{eqnarray}
  \vec{E}^{n}_{\rm mod} &=& \vec{E}^{n} + \frac{2m}{e} \gamma^{n-1/2} q \Omega_0^2 x \vec{e}_x, \\
  \vec{B}^{n}_{\rm mod} &=& \vec{B}^{n} + \frac{2m}{e} \gamma^{n-1/2} \vec{\Omega}_0.
\end{eqnarray}
After the calculation of motion of equations, we could obtain the electric current $\vec{J}^{n+1/2}$ 
by using
\begin{equation}
  \vec{J}^{n+1/2} = \sum_{\rm particle} e \vec{v}^{n+1/2} S(\vec{x}^{n+1/2}),
\end{equation}
where $S(\vec{x})$ is the so-called shape function providing a way the particle
density and velocity are distributed into the grid around its center.

The semi-implicit time integration with the finite-difference method in space 
was used to advance a set of Maxwell's equations.  Our scheme is described by
\begin{eqnarray}
 \frac{1}{c} \left( \frac{\vec{E}^{n+a}-\vec{E}^{n}}{\Delta t} -
\frac{\vec{v}_0}{c} \times   \frac{\vec{B}^{n+a}-\vec{B}^{n}}{\Delta t} 
 \right) &=&
 \nabla \times \left( \alpha \vec{B}^{n+a}+(1-\alpha)\vec{B}^{n} \right) - \frac{4 \pi}{c} \vec{J}^{n+1/2}, 
    \label{Ampere_appendix} \\
 \frac{1}{c} \left( \frac{\vec{B}^{n+a}-\vec{B}^{n}}{\Delta t} \right) &=&
 -\nabla \times \left( \alpha \vec{E}^{n+a}+(1-\alpha)\vec{E}^{n} \right), 
    \label{Faraday_appendix}
\end{eqnarray}
where $\vec{v}_0 = \vec{\Omega}_0 \times \vec{r}$ is the term of a noninertial frame rotating with the
angular velocity $\Omega_0$.  
$\nabla \times$ is replaced by either the finite-difference equation in space or
$i \vec{k} \times$ in Fourier space.  
The semi-implicit parameters of $\alpha$ and $a$ can control the numerical stability.
If $\alpha=1/2$ and $a=1$, the time integration becomes basically 
the centered difference approximation in time.  On the other hand, 
the semi-implicit parameters of $\alpha > 0.5$ and $a<1$ can make 
the time integration numerically stable.

To solve the above Maxwell's equations in the finite difference approximation in space, 
the electric field $E^{n+a}$ is removed from Eqs.(\ref{Ampere_appendix}) and (\ref{Faraday_appendix}),
and we first solve the second-order differential equation for $B^{n+a}$ by the conjugate gradient method.
Then we obtain $E^{n+a}$ from Eq.(\ref{Ampere_appendix}).  Another method is to use Fourier transformation
where the so-called numerical dispersion error is free, but this may present a disadvantage in 
massive parallel computing.

For proceeding to the next time step, we need the information of velocity $\vec{v}^{n+a-1/2}$ and 
position $\vec{x}^{n+a}$, which are obtained by the interpolation of
\begin{eqnarray}
  \vec{v}^{n+a-1/2} &=& a~\vec{v}^{n+1/2}+ (1-a)~\vec{v}^{n-1/2}, \\
  \vec{x}^{n+a} &=& a~\vec{x}^{n+1}+ (1-a)~\vec{x}^{n}.
\end{eqnarray}
In this method, all physical quantities are advanced in time with the time step of $a~\Delta t$, where $a$
is the same value used in the set of  Maxwell's equations.
By using $a < 1$, we can numerically suppress both the high-frequency and the large wave number 
electromagnetic and electrostatic waves.
In our MRI simulation, we used $\alpha = 0.5 \sim 0.51$ and $a=0.98 \sim 1.0$.

In the moment calculation of $\vec{J}$, we used the standard charge conservation method proposed by \citet{Villasenor92}.  
This method, in general, does not require solving the Poisson equation, 
but our open shearing box boundary condition may lead to small but nonnegligible numerical inconsistency 
between the electrostatic field calculated by Eq.(\ref{ampere_eq}) and the Poisson equation of Eq.(\ref{poisson_eq}), 
because the motion of particles across the boundary and the electric and magnetic fields at the boundary
are interpolated to match the open shearing box condition.  
Therefore, to calculate the electrostatic field at the boundary correctly, 
we also solve the Poisson equation at every time step given by
\begin{equation}
  \nabla \cdot \left( \vec{E}^{n+a}- \frac{\vec{v}_0}{c} \times \vec{B}^{n+a} \right)=4 \pi \rho_c^{n+a}.
\end{equation}
The correction of the electric field is useful for suppressing electrostatic waves excited by
numerical errors around the boundary.
\bibliography{mri}

\clearpage
\begin{table}
\begin{center}
\caption{Simulation parameters \label{tbl1}}
\begin{tabular}{c|cccc}
\tableline\tableline
RUN & A & B & C & D \\
\tableline
$\beta$ & 96 & 384 & 1536 & 6144 \\
$(V_A/\Omega_0)/\Delta$ & 50.0 & 35.4 & 25.0 & 17.7 \\
$(v_{t\pm}/\Omega_{c\pm})/\Delta$ & 28.2 & 40.0 & 56.5 & 80.0 \\
$\Omega_{c\pm}/\Omega_0$ & 10.0 & 10.0 & 10.0 & 10.0 \\
$v_{t\pm}/c$ & 0.0707 & 0.1 & 0.141 & 0.2 \\
$V_A/c$ & $1.25 \times 10^{-2}$ & $8.84\times 10^{-3}$ & $6.25\times 10^{-3}$ & $4.42\times 10^{-3}$ \\
$N_x=N_y$ & 700 & 600 & 400 & 300 \\
$L_x=L_y$ & 2.23 & 2.70 & 2.55 & 2.70 \\ 
$N_p/$cell & 400 & 800 & 1600 & 6400 \\
\tableline\tableline
\end{tabular}
\end{center}
\end{table}

\clearpage

\clearpage

\clearpage

\clearpage

\clearpage

\clearpage

\clearpage
\clearpage

\clearpage

\clearpage

\clearpage

\clearpage

\clearpage

\end{document}